\documentclass[12pt]{iopart}
\usepackage[T1]{fontenc}
\usepackage[latin9]{inputenc}
\usepackage{amsbsy}
\usepackage{amstext}
\usepackage{graphicx}

\makeatletter
\def\ssr{\ref@jnl{Space~Sci.~Rev.}}     
\expandafter\let\csname equation*\endcsname\relax 
\expandafter\let\csname endequation*\endcsname\relax
\usepackage{iopams}
\usepackage{setstack}
\usepackage{aas_macros}
\usepackage{microtype}

\usepackage[numbers, sort&compress]{natbib}

\def\newblock{\hskip .11em plus .33em minus .07em}

\newcommand{\eqref}[1]{(\ref{#1})}

\@ifundefined{showcaptionsetup}{}{%
 \PassOptionsToPackage{caption=false}{subfig}}
\usepackage{subfig}
\makeatother

\begin{document}
\global\long\def\g{\nabla}
\global\long\def\E{\boldsymbol{E}}
\global\long\def\i{\mathrm{i}}
\global\long\def\x{\boldsymbol{x}}
\global\long\def\B{\boldsymbol{B}}
\global\long\def\vv{\boldsymbol{v}}
\global\long\def\gv{\g_{v}}
\global\long\def\n{\delta n_{\i}}
\global\long\def\f{\delta f_{\i}}
\global\long\def\de{\delta\E}
\global\long\def\db{\delta\B}
\global\long\def\F{F_{0\i}}
\global\long\def\k{\boldsymbol{k}}
\global\long\def\d{\cdot}
\global\long\def\t{\times}
\global\long\def\kx{k_{\perp}}
\global\long\def\kp{k_{\|}}
\global\long\def\u{\boldsymbol{u}}
\global\long\def\sp{\sigma}
\global\long\def\oo{\infty}
\global\long\def\tpa{T_{0\i\|}}
\global\long\def\tp{T_{0\i\perp}}
\global\long\def\vtpa{v_{\mathrm{th}\i\|}}
\global\long\def\vtp{v_{\mathrm{th}\i\perp}}
\global\long\def\vp{v_{\perp}}
\global\long\def\vpa{v_{\|}}
\global\long\def\dby{\delta B_{y}}
\global\long\def\dbp{\delta B_{\|}}
\global\long\def\dbx{\delta B_{x}}
\global\long\def\c{c_{\mathrm{ref}}}
\global\long\def\kh{\hat{k}}
\global\long\def\kph{\hat{k}_{\|}}
\global\long\def\kxh{\hat{k}_{\perp}}
\global\long\def\vph{\hat{v}_{\perp}}
\global\long\def\vpah{\hat{v}_{\|}}
\global\long\def\Alf{\textrm{Alfvén}}
\global\long\def\oci{\Omega_{\mathrm{ci}}}
\global\long\def\oce{\Omega_{\mathrm{ce}}}
\global\long\def\ocs{\Omega_{\mathrm{c\sp}}}
\global\long\def\va{v_{\mathrm{A}}}
\global\long\def\di{d_{\i}}
\global\long\def\m{m_{\i}}
\global\long\def\e{\mathrm{e}}
\global\long\def\p{\mathrm{p}}
\global\long\def\im{\mathrm{i}}
\global\long\def\vts{v_{\mathrm{th}\sp}}
\global\long\def\vti{v_{\mathrm{th},\mathrm{i}}}
\global\long\def\vte{v_{\mathrm{th},\e}}

\title[Comparative study of gyrokinetic, hybrid-kinetic and fully kinetic
wave physics]{Comparative study of gyrokinetic, hybrid-kinetic and fully kinetic
wave physics for space plasmas }

\author{D~Told$^{1}$, J~Cookmeyer$^{1,2}$, F~Muller$^{3,4}$, P~Astfalk$^{3}$,
F~Jenko$^{1}$}

\address{1) Department of Physics and Astronomy, University of California,
Los Angeles, CA 90095, USA}

\address{2) Haverford College, 370 Lancaster Avenue, Haverford, PA 19041,
USA}

\address{3) Max-Planck-Institut für Plasmaphysik, Boltzmannstr. 2, D-85748
Garching, Germany}

\address{4) ENSTA ParisTech, Université Paris-Saclay, 828 Boulevard des Marechaux,
91762 Palaiseau Cedex - France}

\address{}

\submitto{\NJP}
\begin{abstract}
A set of numerical solvers for the linear dispersion relations of
the gyrokinetic, the hybrid-kinetic, and the fully kinetic model is
employed to study the physics of the kinetic $\Alf$ wave and the
fast magnetosonic mode in these models. In particular, we focus on
parameters that are relevant for solar wind oriented applications
(using a homogeneous, isotropic background), which are characterized
by wave propagation angles averaging close to 90$^{\circ}$. It is
found that the gyrokinetic model, while lacking high-frequency solutions
and cyclotron effects, faithfully reproduces the fully kinetic $\Alf$
wave physics close to, and sometimes significantly beyond, the boundaries
of its range of validity. The hybrid-kinetic model, on the other hand,
is much more complete in terms of high-frequency waves, but owing
to its simple electron model it is found to severely underpredict
wave damping rates even on ion spatial scales across a large range
of parameters, despite containing full kinetic ion physics.
\end{abstract}
\maketitle

\section{Introduction}

Plasmas both in nature or in the laboratory often exist in hot and/or
dilute states, resulting in very long mean free paths of the charged
particles. Under such conditions, traditional fluid approaches like
magnetohydrodynamics are often still applicable to the large-scale
evolution of the system, but they do not account for kinetic effects
like wave-particle interactions. On the one hand, these may cause
microinstabilities resulting in enhanced transport, and on the other
hand, through effects like Landau and cyclotron damping they may determine
how energy is dissipated at the end of the turbulent spectrum. The
former issue, i.e. microinstabilities causing enhanced turbulent transport,
is one of the key challenges in fusion research \cite{Doyle07}, whereas
the latter question about the nature of energy dissipation has recently
been termed ``one of the outstanding open problems in space physics''
\cite{Bruno13}.

In order to describe such phenomena, one would ideally solve the fully
kinetic Vlasov-Maxwell system. Unfortunately, this is usually not
feasible for a realistic system involving multiple scales from the
global system size down to the Debye length scale. In order to circumvent
this problem, often approximations are imposed, either by artificially
reducing the dimensionality of such simulations, or by employing a
reduced model that is computationally less intense. Such theories
have been developed both for laboratory systems such as fusion plasmas,
and for other systems like the solar wind, the corona, or astrophysical
plasmas. While the gyrokinetic model \cite{Brizard07} is a de-facto
standard for turbulence modeling in core fusion plasmas---sometimes
employing further optimizations such as a low mass ratio expansion
for the electron dynamics \cite{Lin01,Holod13}---, in space physics
several different kinetic models are used by various groups to describe
similar physics (see below for references). A recent publication \cite{Parashar15}
has sparked the ``turbulent dissipation challenge'' (similar in
spirit to the GEM reconnection challenge \cite{GEM01}), which aims
to compare the various reduced models to determine their capability
of modeling various aspects of plasma physics that are relevant to
the solar wind, but which should also prove instructive for neighboring
fields. The present work aims to aid this effort by comparing, within
a linear framework, the gyrokinetic model \cite{Howes06,Rogers07,Howes08,Howes11,TenBarge2013,Pueschel14,Kobayashi14,Told15},
a widely used hybrid-kinetic model \cite{Chodura75,Sgro76,Harned82,Winske85,Brecht88,Swift96,Winske03,Gargate07,Valentini07,Mueller11,Kempf13,Cheng13,Kunz14},
and a fully kinetic ion and electron description.

To date, such comparisons have been mainly performed between two models
\cite{Howes06,Tronci15}, but so far there exists, for instance, no
direct comparison between the gyrokinetic and the hybrid-kinetic model
for parameters relevant to the solar wind, a gap that we intend to
fill with the present work. This paper is structured as follows: In
Sec.~\ref{sec:Models}, we give a brief account of the three models
that will be compared in the present work. In the main part, Sec.~\ref{sec:Results},
a detailed comparison between these three models is performed for
two commonly analyzed waves, the kinetic $\Alf$ wave, and the fast
magnetosonic mode/whistler. In Sec.~\ref{sec:conclusions}, we summarize
these results and provide some conclusions as well as an outlook.

\section{Three  kinetic models\label{sec:Models}}

\subsection{The gyrokinetic system of equations}

The gyrokinetic model has originated and been employed successfully
for several decades in the fusion community, and its range of applications
has only recently been extended to space plasmas. Gyrokinetics (GK)
is an analytical limit of full kinetics which is intended for strongly
magnetized plasmas, where the particle motion perpendicular to a strong
magnetic guide field can be expanded in terms of a fast gyration around
that field and a drift motion perpendicular to that field. Several
ordering assumptions must be obeyed in order for this theory to be
valid \cite{Howes06,Brizard07}, i.e. 
\begin{equation}
\frac{\kp}{\kx}\sim\frac{\omega}{\ocs}\sim\frac{c}{\vts}\frac{\delta E_{\perp}}{B}\sim\frac{\delta B}{B}\sim\epsilon,\label{eq:GK}
\end{equation}
where $\epsilon$ is a small parameter. As a consequence of the above
ordering relations, the wave propagation angles that can be described
within GK are constrained to be almost perpendicular to the background
field. In addition, wave frequencies formally need to be much smaller
than the cyclotron frequency of the involved species. In the main
part of this paper, we will analyze how much the wave physics is actually
altered when one or more of these conditions is violated. 

The gyrokinetic dispersion relation for a homogeneous plasma with
isotropic Maxwellian background distribution has been derived in Ref.~\cite{Howes06}.
Here, we use a slightly modified equation compared to their Eq.~(41),
since the derivation of that formula involves a multiplication with
the function $A$ (defined below). However, this function can become
zero for certain complex frequencies, and multiplying the dispersion
relation by $A$ thus introduces spurious solutions. Instead of solving
their Eq.~(41), we directly set to zero the determinant of the matrix
in their Eq.~(C15), \begin{equation}\det\left(\matrix{A&A-B&C\cr
A-B&A-B-\alpha_i/\bar\omega^2&C+E\cr C&C+E&D-2/\beta_i}\right)=0,\label{eq:DR}\end{equation}where 
\begin{eqnarray*}
A & = & \left(1+\frac{T_{0\i}}{T_{0\e}}\right)+\sum_{\sp}\frac{T_{0\i}}{T_{0\sp}}\Gamma_{0}\left(\alpha_{\sp}\right)\xi_{\sp}Z\left(\xi_{\sp}\right)\\
B & = & \left(1+\frac{T_{0\i}}{T_{0\e}}\right)-\sum_{\sp}\frac{T_{0\i}}{T_{0\sp}}\Gamma_{0}\left(\alpha_{\sp}\right)\\
C & = & \frac{1}{e}\sum_{\sp}q_{\sp}\Gamma_{1}\left(\alpha_{\sp}\right)\xi_{\sp}Z\left(\xi_{\sp}\right)\\
D & = & 2\sum_{\sp}\frac{T_{0\sp}}{T_{0\i}}\Gamma_{1}\left(\alpha_{\sp}\right)\xi_{\sp}Z\left(\xi_{\sp}\right)\\
E & = & \frac{1}{e}\sum_{\sp}q_{\sp}\Gamma_{1}\left(\alpha_{\sp}\right),
\end{eqnarray*}
using the same notation as in Ref.~\cite{Howes06}. A Python program
is used to solve for the complex frequencies fulfilling this dispersion
relation, with very similar algorithms as in the recently introduced
HYDROS code for the hybrid-kinetic system (see next section).

\subsection{The hybrid-kinetic system of equations}

The second model that will be examined in this work, is the hybrid
kinetic-ion/fluid-electron model (which we will simply call ``hybrid-kinetic''
in the following). The equations of this model are obtained by taking
a nonrelativistic limit of the fully kinetic equations, and taking
the electron mass to zero. Retaining only a singly charged ion species,
such that $n_{\i}=n_{\e}$, then leads to a system of equations consisting
of the ion Vlasov equation,
\begin{equation}
\frac{\partial f_{\i}}{\partial t}+\vv\cdot\g f_{\i}+\left[\frac{e}{m_{\i}}\left(\E+\frac{\vv\times\B}{c}\right)\right]\cdot\gv f_{\i}=0,\label{eq:Vl}
\end{equation}
and an Ohm's law determining the electric field, which reads
\begin{equation}
n_{\e}\E=-\frac{1}{c}n_{\i}\u_{\i}\times\B+\frac{1}{ce}\boldsymbol{j}\times\B-\frac{1}{e}\nabla P_{\e}+n_{\e}\eta\boldsymbol{j},\label{eq:Ohm}
\end{equation}
with $n_{\i}\u_{\i}=\int_{\mathcal{V}}\vv f_{\i}d^{3}v$, the resistivity
$\eta$, and the electron pressure gradient $\nabla P_{\e}=C\g n_{\e}^{\gamma}$.
The electromagnetic fields are constrained by Faraday's law
\begin{equation}
\frac{\partial\B}{\partial t}=-c\g\times\E,\label{eq:Faraday}
\end{equation}
and the pre-Maxwell version of Ampere's law (which, due to the absence
of the displacement current, implies quasi-neutrality)
\begin{equation}
\g\t\B=\frac{4\pi}{c}\boldsymbol{j}.\label{eq:amp}
\end{equation}
These equations contain the full kinetic ion physics including wave-particle
interactions such as Landau and transit-time damping, as well as cyclotron
resonances. On the other hand, electrons appear only as a neutralizing,
massless background. The equation of state for the electron pressure
gradient is determined by $P_{\e}=Cn_{\e}^{\gamma}$, with $C=n_{\e}^{1-\gamma}T_{0\e}$
and, in this work, $\gamma=1$, corresponding to an isothermal equation
of state. 

A dispersion relation for this kind of model, valid for arbitrary
propagation angle and bi-Maxwellian plasmas, has recently been derived
\cite{Told16aarX}, and the numerical dispersion solver HYDROS (previously
employed in Ref.~\cite{Cerri16}) will be used in this work to obtain
the wave solutions for that model.

\subsection{The fully kinetic system of equations}

Finally, the reference for this comparative analysis is provided by
the nonrelativistic, fully kinetic system of equations, with one Vlasov
equation (Eq.~\ref{eq:Vl}) for each species $\sp$, and the full
Maxwell equations including the displacement current
\begin{eqnarray*}
\nabla\d\boldsymbol{E}=4\pi\sum_{\sp}q_{\sp}\int_{\mathcal{V}}f_{\sp}d^{3}v\\
\nabla\d\boldsymbol{B}=0\\
\\
\frac{\partial\B}{\partial t}=-c\g\times\E\\
\g\t\B=\frac{4\pi}{c}\boldsymbol{j}+\frac{1}{c}\frac{\partial\boldsymbol{E}}{\partial t},
\end{eqnarray*}
with $\boldsymbol{j}=\sum_{\sp}q_{\sp}\int_{\mathcal{V}}\boldsymbol{v}f_{\sp}d^{3}v$,
and the integral running over the full velocity space $\mathcal{V}$.
The inclusion of the displacement current in Ampére's law introduces
(in normalized units) a new dimensionless parameter $\va/c$ (with
$\va=B/\sqrt{4\pi m_{\i}n_{\i}}$), which is set to $0.01$ throughout
this work, small enough to avoid discrepancies due to quasineutrality
violations (in the solar wind, e.g. from Ref.~\cite{Sahraoui10},
$\va/c\approx2\cdot10^{-4}$). The fully kinetic wave solutions are
obtained here using the DSHARK dispersion relation solver \cite{Astfalk15}
in the limit of an isotropic Maxwellian background distribution.

\section{Comparative study of linear wave physics\label{sec:Results}}

In the main part of the paper, a comparison of gyrokinetic (GK), hybrid-kinetic
(HK), and fully kinetic (FK) wave physics is provided for some of
the waves that are encountered in conditions found in the solar wind
or magnetospheric plasmas, and which have been the center of various
previous studies, namely the kinetic $\Alf$ wave (KAW), and the fast
magnetosonic mode/whistler.

\subsection{On the choice of physical parameters}

For the analysis detailed in this section, the focus is on parameters
that are suitable for magnetized plasmas, specifically for the small
(kinetic) scales close to or below the ion gyroradius ($k\rho_{\i}\sim1$,
where $\rho_{\i}=m_{\i}\vti c/eB$, and $\vti=\sqrt{2T_{\i}/\m}$
is the ion thermal velocity\footnote{Note that this definition agrees with the one of Ref.~\cite{Howes06},
but differs by a factor $\sqrt{2}$ from the definition used in Ref.~\cite{Told15}. }.) or ion skin depth scale ($kd_{\i}\sim1$, where $d_{\i}=c/\omega_{\mathrm{pi}}$),
where kinetic effects become significant. Owing to the nature of the
MHD cascade \cite{GS95,Boldyrev06}, if there is a sufficiently broad
spectral range between an approximately isotropic injection scale
and the scale of interest, the turbulent fluctuations in said range
of interest will exhibit anisotropic wavevectors with $\kp\ll\kx$,
low frequency $\omega\ll\oci$, and small amplitude $\delta B\ll B_{0}$. 

These are the requirements for the validity of gyrokinetic theory
(see Eq.~\ref{eq:GK}), and both theoretical arguments and spacecraft
observations \cite{Howes08a,Sahraoui10,Chen10,Chen13a} indicate that
these requirements are fulfilled in the kinetic wavenumber range of
the solar wind close to 1\,AU, although other observations point
to the occurrence of frequencies significantly larger than $\oci$
\cite{Narita11}. In this work, we will remain agnostic toward these
questions, and will compare the three different models both in regimes
that fulfill the above requirements as well as ones that do not. 

Most significantly, we emphasize that all wavenumber spectra in this
work are created using a constant propagation angle, which is most
likely \emph{not} realized in actual plasmas. In fact, following the
critical balance arguments by Goldreich/Sridhar (GS) \cite{GS95},
but also Boldyrev \cite{Boldyrev06}, MHD turbulence is characterized
by $\kp\propto\kx^{\alpha}$, i.e. the ratio between $\kp$ and $\kx$,
and thus the propagation angle, is not independent of the wavenumber.
The exponent $\alpha$ takes a value of $2/3$ for GS, and $1/2\le\alpha\le2/3$
for Boldyrev, and can assume even smaller values (1/3) in the kinetic
range, or for turbulence weakened by dissipation \cite{Howes11a}.
In the present work, however, our aim is not to reproduce realistic
spectra, but to compare the three different models, justifying the
choice of a simple, fixed, $\kp/\kx$ ratio. 

Spacecraft observations from Refs.~\cite{Sahraoui10,Narita11} indicate
that the average propagation angle in the solar wind is 87.8/87.7$^{\circ}$,
respectively, with deviations of at most $\sim15{}^{\circ}$ \cite{Sahraoui10}
($\sim30{}^{\circ}$ \cite{Narita11}). Therefore, most tests in the
following will be performed at angles close to the average value,
although a more extreme propagation angle of 60$^{\circ}$ will be
analyzed as well. 

Finally, we note that the solar wind plasma usually exhibits temperatures
that are anisotropic with respect to the mean magnetic field (see,
e.g., Ref.~\cite{Bale09}). Here, we focus on an isotropic background
and leave a comparison of anisotropy driven instabilities (such as
the mirror and the firehose mode) for future work. Previous work indicates
that standard gyrokinetics should be able to reproduce the behavior
of the mirror mode for almost perpendicular propagation, but may require
generalizations to treat the firehose instability \cite{Porazik13}.
The GK dispersion relation used here \cite{Howes06} is formulated
for isotropic background temperature, and would thus have to be generalized
for a future comparison.

\subsection{Kinetic $\protect\Alf$ waves}

The first test case for which the three candidate models will be compared
is the kinetic $\Alf$ wave, i.e. the continuation of the MHD shear
$\Alf$ wave into the range where $k\rho_{\i}\sim1$ (where FLR effects
become important) or $kd_{\i}\sim1$, where ion inertia becomes significant
and the electron and ion motion thus decouple.

\subsubsection{Propagation angle dependence\label{sub:Kawprop}}

First, a set of fixed wave propagation angles is chosen, and scans
over the magnitude of the wave vector are performed in order to obtain
a dispersion relation of the KAW for this parameter set. For this
analysis, $\beta_{\i}=\beta_{\e}=1$ is chosen (with $\beta_{\sp}=8\pi n_{\sigma}T_{\sigma}/B^{2}$),
and wavenumbers are scanned from $kd_{\i}=k\rho_{\i}=0.01$ to $kd_{\i}=20$. 

\begin{figure}
\subfloat[]{\includegraphics[width=0.5\textwidth]{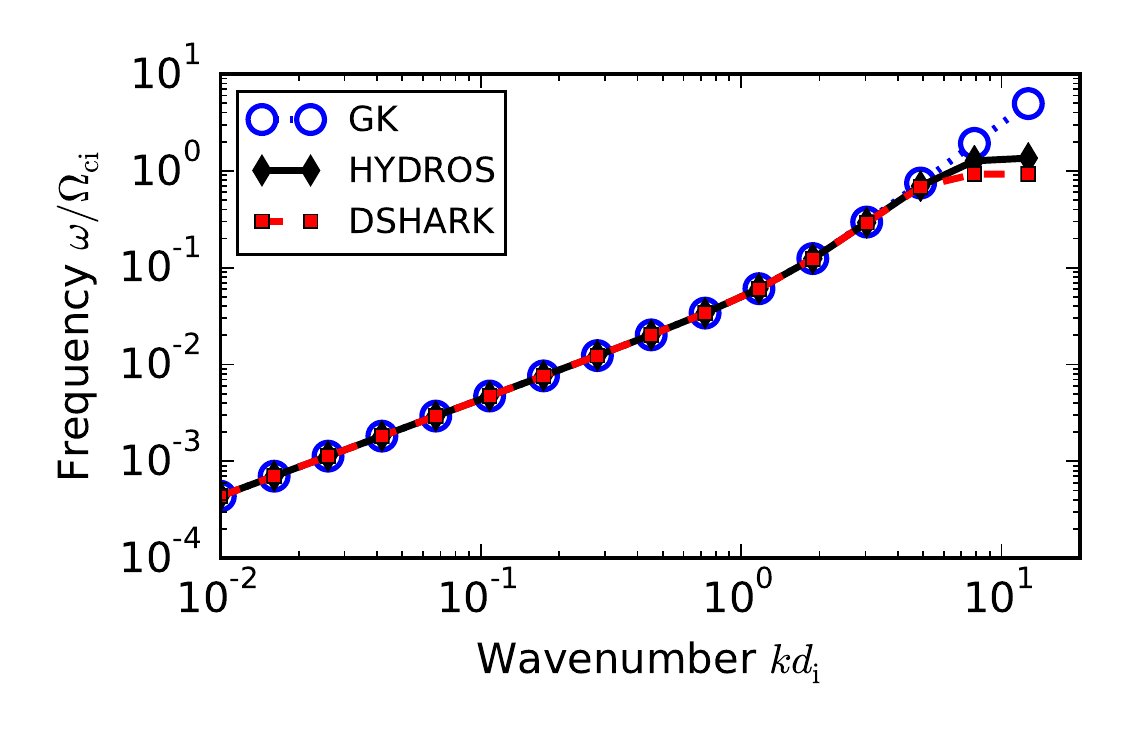}

}\subfloat[]{\includegraphics[width=0.5\textwidth]{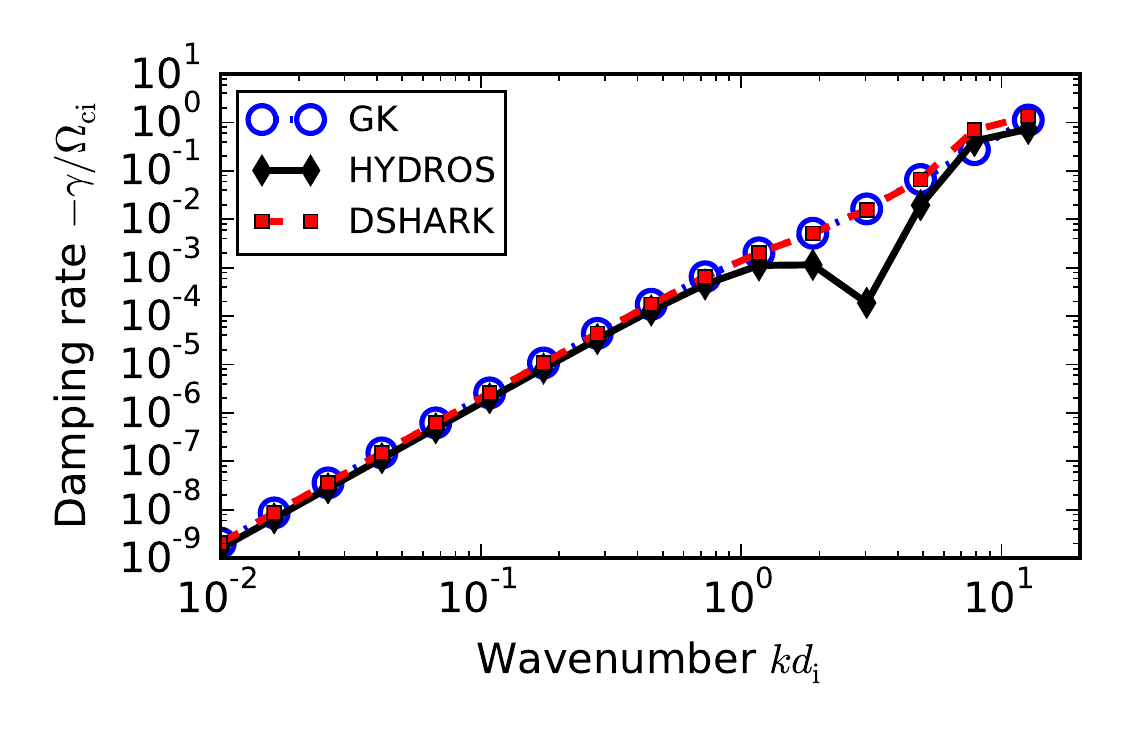}

}

\caption{Wavenumber scan for the kinetic $\protect\Alf$ wave, for $\beta_{\protect\i}=\beta_{\protect\e}=1$
and a fixed propagation angle $\theta=87.5^{\circ}$. Plot a) shows
the frequencies, and plot b) compares the damping rates. Blue circles
denote the gyrokinetic results, black diamonds mark the hybrid-kinetic
HYDROS points, and red squares are used to plot the fully kinetic
DSHARK results. \label{fig:KAWtheta1}}
\end{figure}

Let us analyze the results presented in Fig.~\ref{fig:KAWtheta1}
first. For this figure, we set $\theta=87.5^{\circ}$. As is apparent
from the left hand part a) of the figure, the frequencies for all
models agree very well over most of the wavenumber range, $0.01\leq kd_{\i}\lesssim6$,
until the KAW reaches the cyclotron frequency. At this point, the
gyrokinetic model, lacking any physics effects related to the cyclotron
motion, continues to exhibit an increasing frequency, whereas the
frequencies from both the hybrid-kinetic and the fully kinetic code
roll over (though not exactly at the same value) and asymptote against
a frequency close to the ion cyclotron frequency $\oci$. 

In the right hand part, Fig.~\ref{fig:KAWtheta1}b), the corresponding
damping rates are presented. Here, relatively good agreement between
all models is observed in the wavenumber range $0.01\leq kd_{\i}\leq1$.
While the GK model agrees very well with full kinetics in that range,
the hybrid-kinetic model underestimates the damping rates there by
about 25\%. This discrepancy is the first hint of a common theme that
will be explored more thoroughly in the following sections, namely
the absence of electron Landau and transit time damping in the hybrid-kinetic
model. Importantly, and perhaps counterintuitively for a model that
includes full kinetic ion physics, this can and will lead to discrepancies
on \emph{ion} spatial scales. 

Here, this discrepancy becomes more obvious as the wavenumber is increased
to $kd_{\i}\gtrsim1$, where the gap between the HYDROS and DSHARK
damping rates widens to about three orders of magnitude, until cyclotron
damping becomes significant at $kd_{\i}\sim6$, closing the gap in
damping rates and reestablishing agreement between the hybrid and
fully kinetic physics. GK, on the other hand, missing the cyclotron
damping effect, does not consistently agree with full kinetics in
that range. 

\begin{figure}
\subfloat[]{\includegraphics[width=0.5\textwidth]{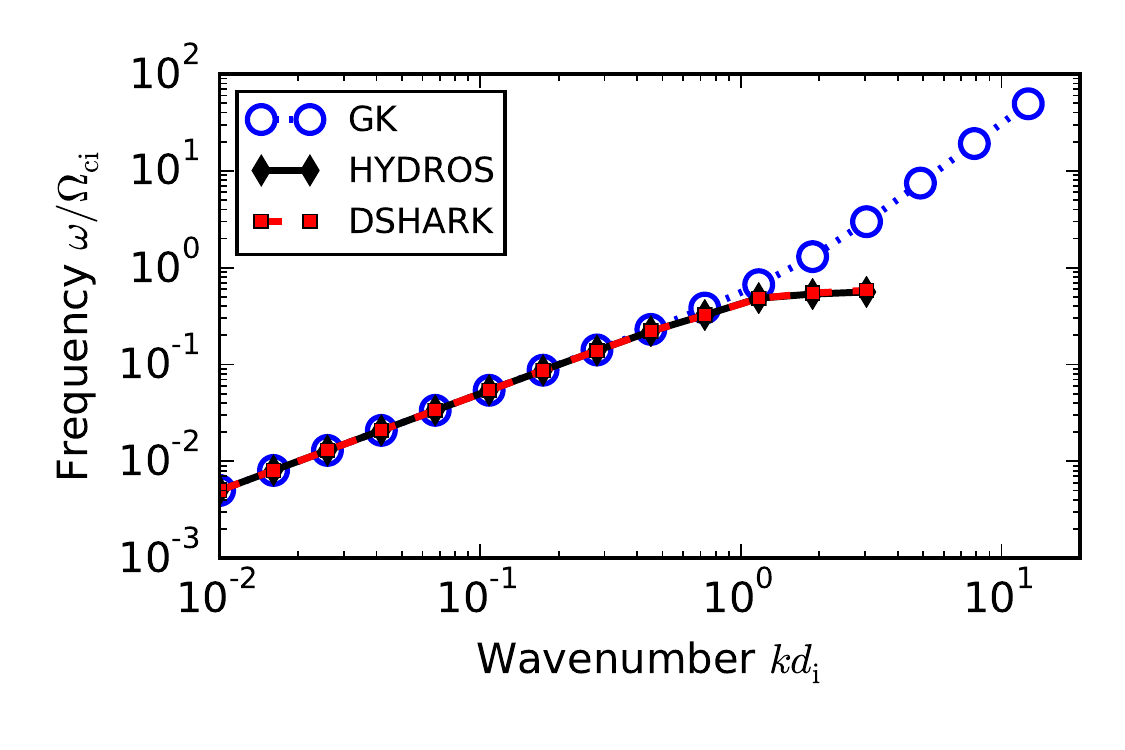}

}\subfloat[]{\includegraphics[width=0.5\textwidth]{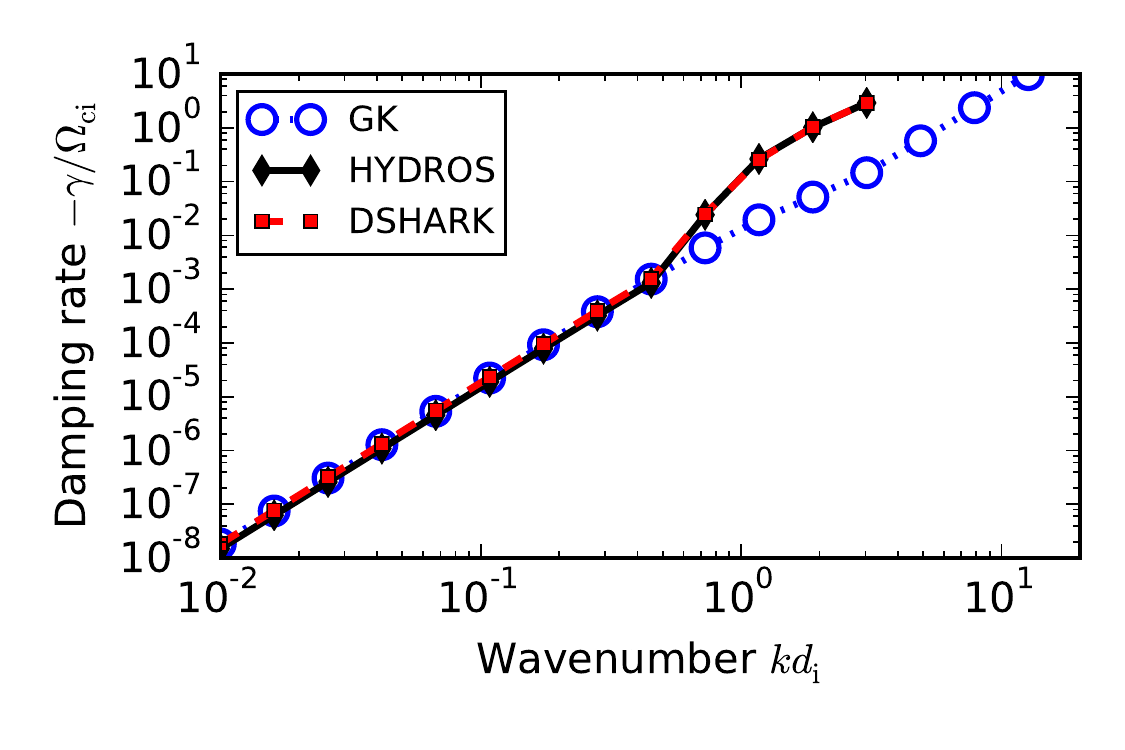}

}

\caption{Wavenumber scan for the kinetic $\protect\Alf$ wave, for $\beta_{\protect\i}=\beta_{\protect\e}=1$
and a fixed propagation angle $\theta=60^{\circ}$. Plot a) shows
the frequencies, and plot b) compares the damping rates.\label{fig:KAWtheta2}}
\end{figure}

Before examining such discrepancies in more detail, the linear KAW
dispersion relation is studied for one more propagation angle, namely
$\theta=60^{\circ}.$ The results for this scan are shown in Fig.~\ref{fig:KAWtheta2}.
As before, very good agreement of all frequencies is observed at low
wavenumbers, until the wave frequency approaches the ion cyclotron
frequency at $kd_{\i}\approx0.8$. At higher $k$, the frequencies
in the hybrid-kinetic and fully kinetic solver roll over, while in
GK they pass through the cyclotron frequency and continue with an
$\omega\propto k^{2}$ relation. 

For this propagation angle, the low-$k$ damping rates between all
models agree very well, indicating that ion Landau damping is dominant
for these parameters. At $kd_{\i}\approx0.6$, the damping rates returned
by the GK solver start to deviate from those of HYDROS and DSHARK,
due to the missing cyclotron damping. 

However, it is noteworthy that any agreement is found at all between
GK and the other models for this propagation angle, considering that
the ordering $\kp\ll\kx$ is a fundamental requirement of gyrokinetics.
For $\theta=60^{\circ}$, however, $\kp\approx0.58\kx$, technically
violating the aforementioned ordering. This finding (if found to be
independent of physics parameters such as $\beta_{\i}$ and $\beta_{\e}$),
is of relevance to systems like the solar wind considering the limited
variability of the propagation angles found in the solar wind \cite{Sahraoui10,Narita11}.
Since GK can apparently still predict useful KAW wave frequencies
and damping rates for propagation angles as small as $60^{\circ}$,
we may expect that, if the model fails in systems like the solar wind,
it is more likely to do so because of the lack of fast waves and cyclotron
effects, not because of the $\kp$ ordering. 

\subsubsection{Beta dependence\label{sub:KAWbeta}}

In this section, we compare the GK, hybrid-kinetic and fully kinetic
solvers for different beta values. A fixed propagation angle of $\theta=85^{\circ}$
is chosen (i.e. slightly less oblique than the average observed angle
in the solar wind), and wavenumber spectra for two separate $\beta_{\i}=\beta_{\e}$
values are analyzed. 

\begin{figure}
\subfloat[]{\includegraphics[width=0.5\textwidth]{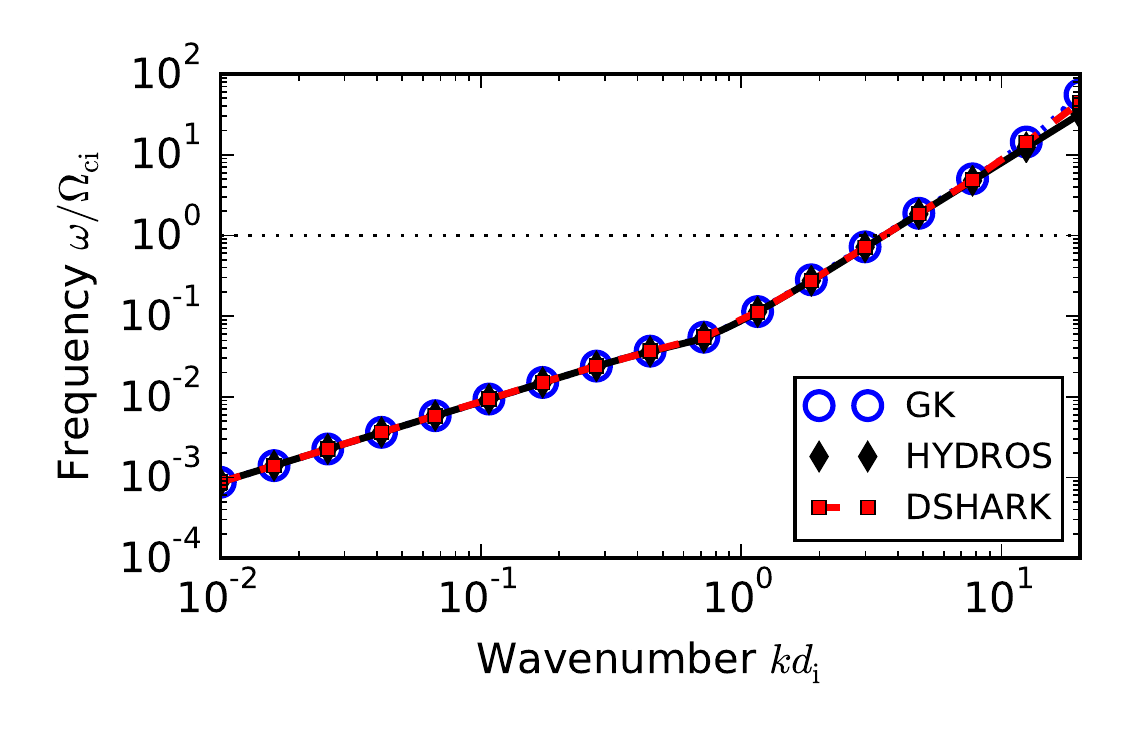}

}\subfloat[]{\includegraphics[width=0.5\textwidth]{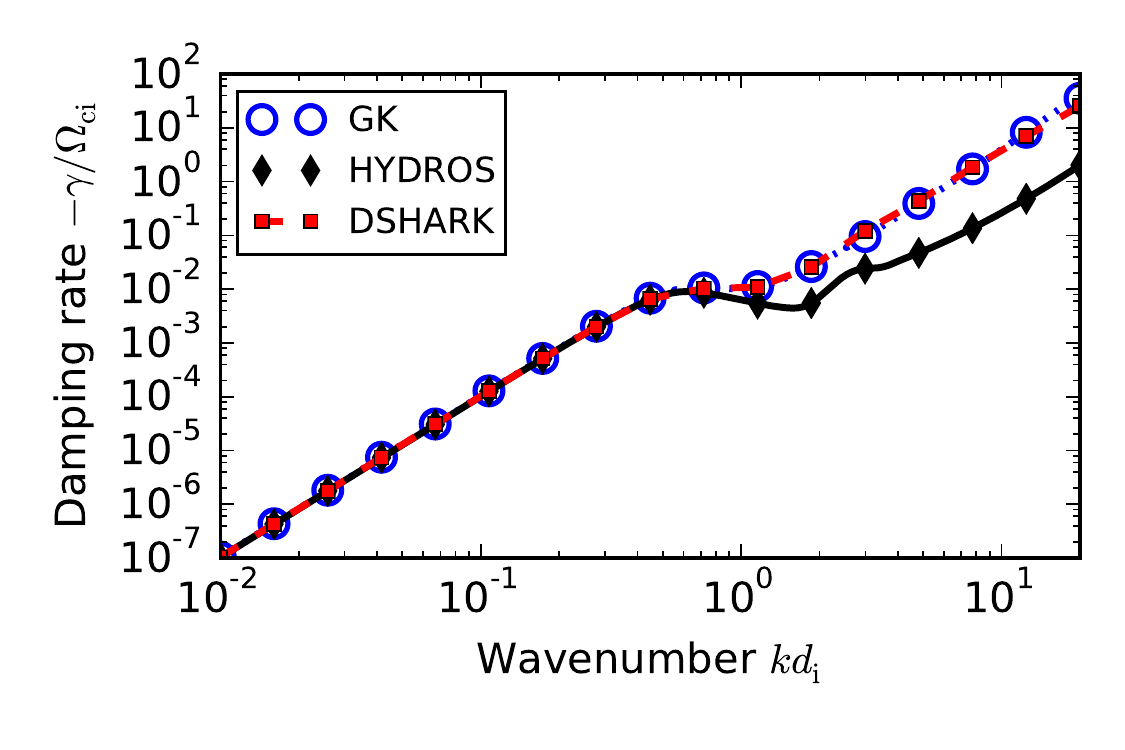}

}

\caption{Wavenumber spectra for GK, hybrid-kinetic and fully kinetic descriptions
of the kinetic $\protect\Alf$ wave. Parameters are $\beta_{\protect\i}=\beta_{\protect\e}=5$,
and $\theta=85^{\circ}$.\label{fig:KAWbeta5}}
\end{figure}

First, let us examine the plots shown in Fig.~\ref{fig:KAWbeta5},
which were obtained for $\beta_{\i}=\beta_{\e}=5$. Here, remarkably,
very good agreement of wave frequencies is found for all models, across
the complete wavenumber range. In this case, the KAW has a right-handed
polarization and is thus not affected by the ion cyclotron resonance.
The GK dispersion relation thus remains valid significantly beyond
the ion cyclotron frequency. This behavior of the KAW for strongly
oblique propagation has been both numerically observed and analytically
derived before, and is discussed in Refs.~\cite{Gary86,Boldyrev13}. 

Even more surprising is the comparison of the damping rate curves
shown in Fig.~\ref{fig:KAWbeta5}b), where the GK model is found
to reproduce the KAW damping rates more accurately than the hybrid-kinetic
model. At low wavenumbers $kd_{\i}\lesssim0.7$, all three models
agree very well. At larger wavenumbers, though, electron Landau damping
appears to become significant. Collisionless damping mechanisms involving
the electron species are, however, completely absent from the hybrid-kinetic
model, resulting in damping rates (potentially caused by anomalous
ion cyclotron damping \cite{Kadomtsev68}) that are about one order
of magnitude smaller than those of GK or full kinetics in the range
$kd_{\i}\gtrsim1$.

These circumstances need to be interpreted cautiously, however: while
for these parameters the representation of the KAW is indeed more
accurate in GK than in hybrid-kinetic, this may not automatically
carry over to a turbulent state, where a KAW cascade may transfer
energy into other waves that occur close to the cyclotron frequency,
like ion Bernstein modes. The latter type of wave is not contained
in gyrokinetics, and such effects are thus absent. The question about
how important those effects are, cannot be answered in the linear
framework of this study, and will have to be addressed in nonlinear,
turbulent simulations. 

As discussed in Section~\ref{sub:Kawprop} (for a propagation angle
of $87.5^{\circ}$), for $\beta_{\i}=\beta_{\e}=1$ the situation
is qualitatively different, as the KAW wave is left hand polarized,
thus enabling the ion cyclotron resonance, and leading to results
similar to the ones shown in Fig.~\ref{fig:KAWtheta1}. For this
lower $\beta$, we also begin to observe a discrepancy of the hybrid-kinetic
damping rates compared to the other two models, on ion spatial scales. 

\begin{figure}
\subfloat[]{\includegraphics[width=0.5\textwidth]{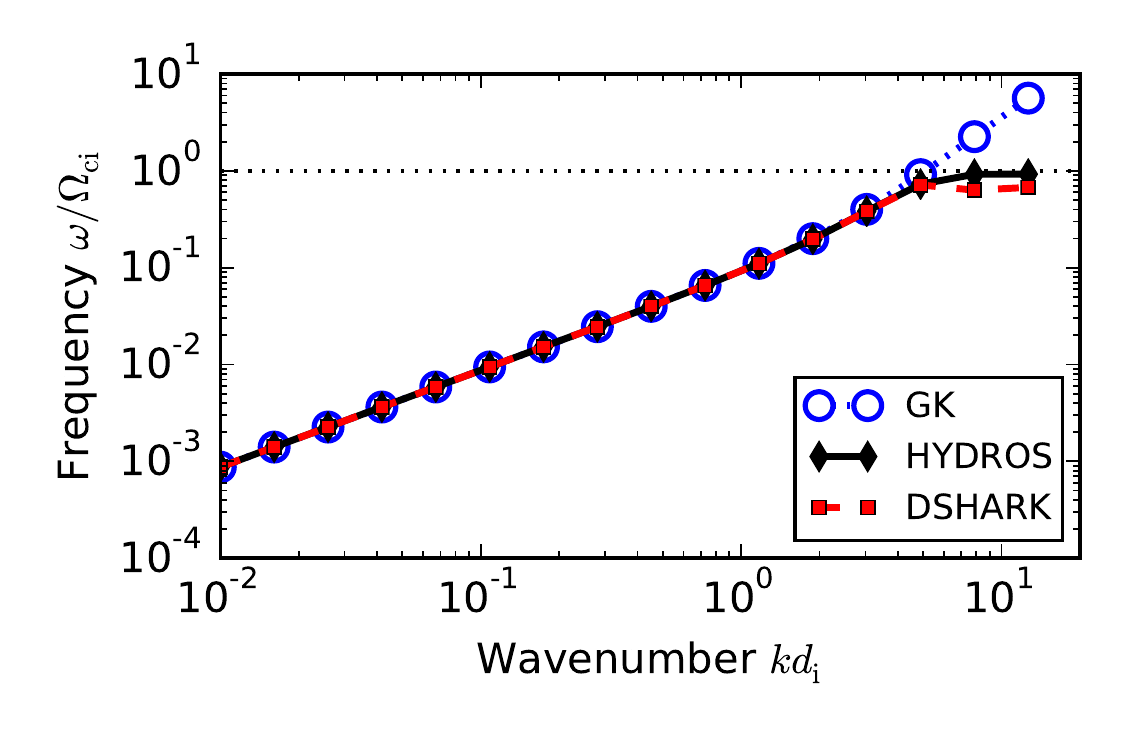}

}\subfloat[]{\includegraphics[width=0.5\textwidth]{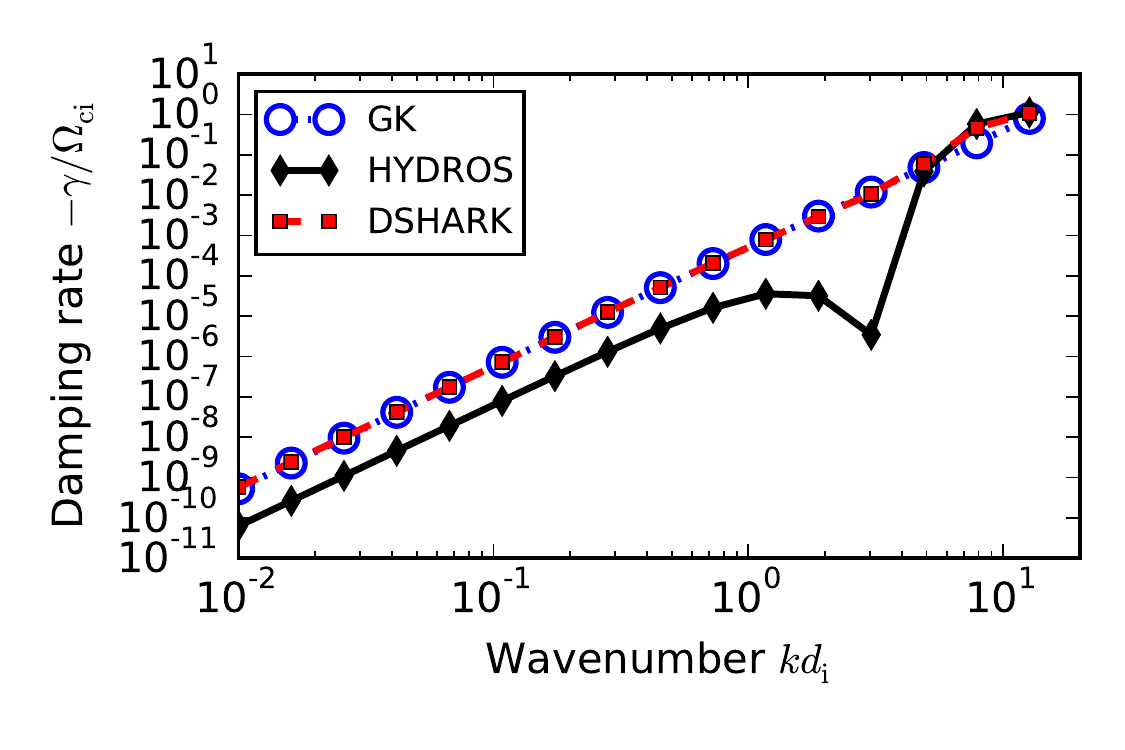}

}

\caption{Wavenumber spectra for GK, hybrid-kinetic and fully kinetic descriptions
of the kinetic $\protect\Alf$ wave. Parameters are $\beta_{\protect\i}=\beta_{\protect\e}=0.2$,
and $\theta=85^{\circ}$.\label{fig:KAWbeta02}}
\end{figure}

Decreasing $\beta$ further to $\beta_{\i}=\beta_{\e}=0.2$, we obtain
the dispersion relations presented in Fig.~\ref{fig:KAWbeta02}.
For the wave frequencies, the picture is qualitatively identical to
that of Fig.~\ref{fig:KAWtheta1} ($\beta=1$, $\theta=87.5^{\circ}$).
The damping rate curves of Fig.~\ref{fig:KAWbeta02}b), however,
reveal a more severe discrepancy between the hybrid-kinetic and the
two other models: Across the whole ion range, $0.01\leq kd_{\i}\lesssim1$,
the hybrid-kinetic model underestimates the KAW damping rate by roughly
one order of magnitude. At higher wavenumbers, where the ion Landau
damping diminishes, this gap widens further, until ion cyclotron damping
becomes dominant at $kd_{\i}\approx4.$

\begin{figure}
\subfloat[]{\includegraphics[width=0.5\textwidth]{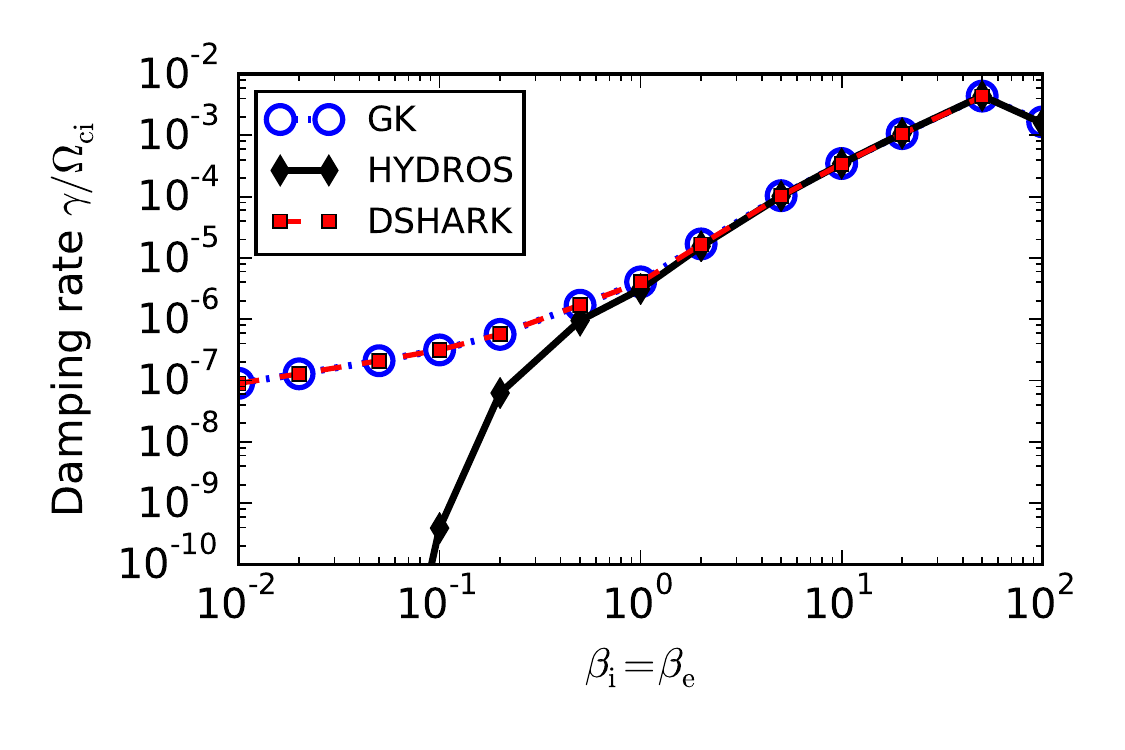}

}\subfloat[]{\includegraphics[width=0.5\textwidth]{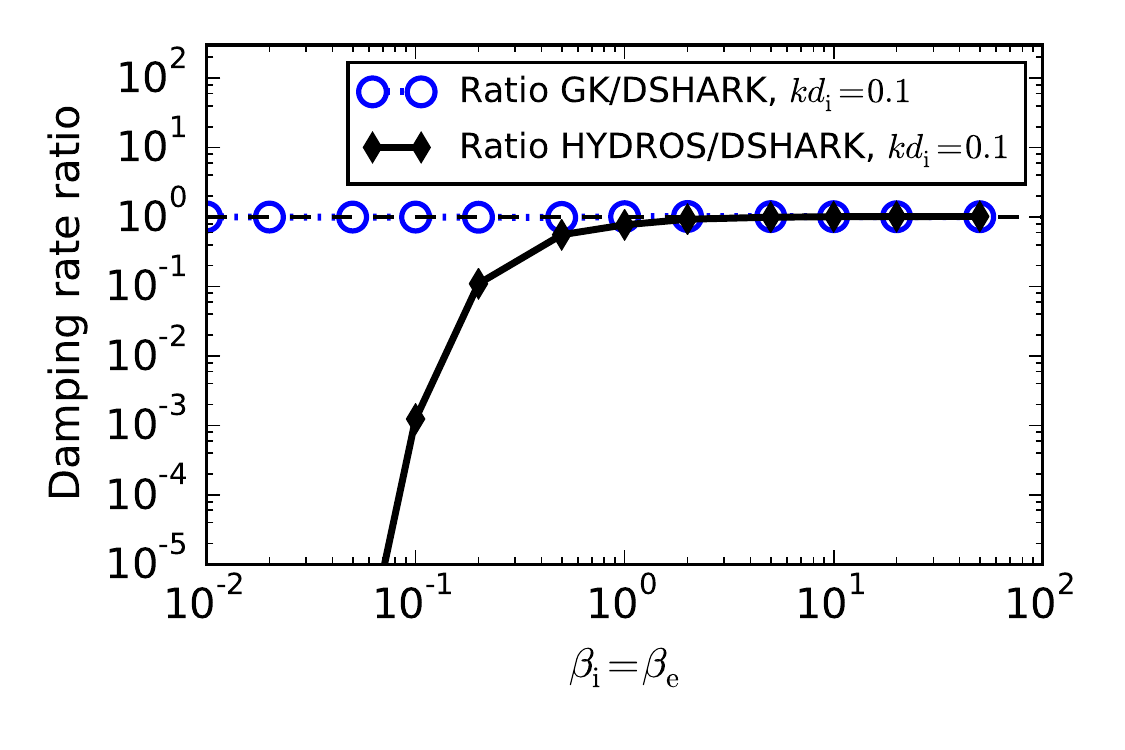}

}

\caption{a) Damping rates in GK, HYDROS and DSHARK for the kinetic $\protect\Alf$
wave, at an ion scale wavenumber of $kd_{\protect\i}=0.1$, for varying
$\beta_{\protect\i}$. b) Ratios GK/hybrid-kinetic and fully kinetic
damping rates. For $\beta_{\protect\i}\lesssim1$, the damping is
dominated by electron Landau damping, resulting in severely underpredicted
damping rates in the hybrid-kinetic model. \label{fig:ratiosHYDROSDSHARK}}

\end{figure}

Since the discrepancy in the damping rates clearly depends on the
beta parameter, another scan is performed in order to characterize
this effect more stringently, this time taking a fixed wavenumber
of $kd_{\i}=0.1$, while varying $\beta_{\i}=\beta_{\e}$ in the range
$0.01\leq\beta_{\i}=\beta_{e}\leq50$. The wavenumber $kd_{\i}$ is
deliberately chosen to lie in the ion range of spatial scales, since
this is the range where one would naively expect a model with fully
kinetic ion treatment to agree with a fully kinetic model. The damping
rates from this scan are shown in Fig.~\ref{fig:ratiosHYDROSDSHARK}a),
and their ratios from the reduced models compared to the fully kinetic
result are plotted in Fig.~\ref{fig:ratiosHYDROSDSHARK}b). For these
parameters, GK agrees very well with full kinetics across the whole
beta range. On the other hand, consistent with the picture that emerged
above, the hybrid-kinetic model shows excellent agreement at high
$\beta\gtrsim2$, while for lower beta values a discrepancy arises.
This discrepancy is about 25\% for $\beta=1$, about one order of
magnitude for $\beta=0.2$, and three orders of magnitude for $\beta=0.1$,
and increases quickly for lower $\beta$.

A physical interpretation of this discrepancy may be obtained by considering
that for an $\Alf$ wave in this range $\omega\approx\kp\va$, where
\[
\va=\frac{B}{\sqrt{4\pi m_{\i}n_{\i}}}=\frac{\vti}{\sqrt{\beta_{\i}}}.
\]
The ion Landau resonance occurs for ions traveling at speeds similar
to the propagation velocity of the $\Alf$ wave, i.e. $\vpa\approx\va$.
For a Maxwellian distributed plasma, such particles are most abundant
for $\beta_{\i}$ values such that $\va\lesssim\vti$, i.e. the ion
Landau resonance is most effective for $\beta_{\i}\gtrsim1$, in agreement
with Fig.~\ref{fig:ratiosHYDROSDSHARK}. On the other hand, for $\beta_{\i}\ll1$
the $\Alf$ wave travels at a velocity faster than most ions and thus
detunes from the ion Landau resonance, resulting in the strongly reduced
damping rates observed in the hybrid model. 

In the case of electron Landau damping, the above statements apply
in a similar way, except that $\va\lesssim\vte$ is now the relevant
condition. Since in a hydrogen plasma the electron thermal velocity
is roughly a factor 43 faster than the ions', this condition is fulfilled
for a much broader range of $\beta$ values, making the electron Landau
damping more resilient against variations of this parameter. For the
examined $\beta$ values, it is thus the electron Landau damping that
remains once the ion Landau damping is removed and that explains the
discrepancy between the hybrid-kinetic and the fully kinetic model.

\subsubsection{Mass ratio effect}

Motivated by the above results, we examine in this section the consequences
of a reduced mass ratio on the behavior of a kinetic $\Alf$ wave.
This question is of interest, as fully kinetic PIC or Eulerian turbulence
simulations are often extremely challenging in a computational sense,
enforcing either a reduced dimensionality of the simulations, or the
removal of physical scale separations such as those between ions and
electrons. The latter is achieved by reducing the mass ratio between
the species, leading to a compression of their spatiotemporal scales. 

\begin{figure}
\subfloat[]{\includegraphics[width=0.5\textwidth]{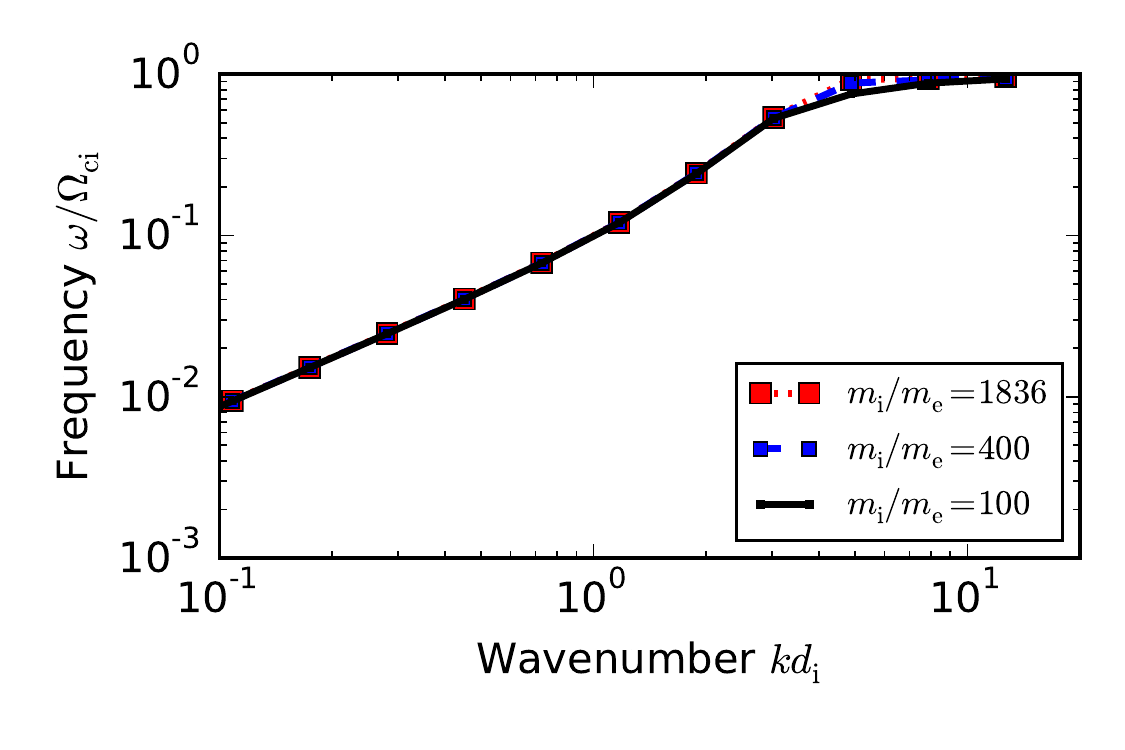}

}\subfloat[]{\includegraphics[width=0.5\textwidth]{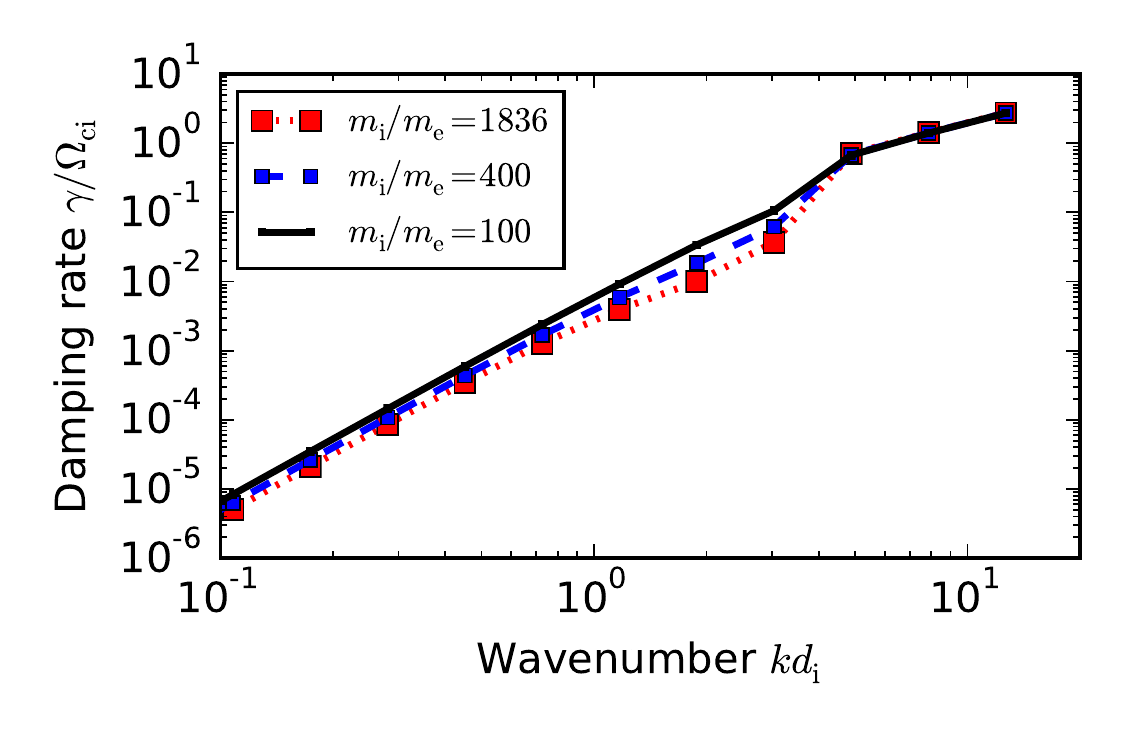}

}

\caption{a) Frequencies and b) damping rates for a kinetic $\protect\Alf$
wave with real proton/electron and reduced mass ratios, produced with
the DSHARK code, using $\beta_{\protect\i}=\beta_{\protect\e}=1$
and a propagation angle of $85^{\circ}$.\label{fig:KAWmass}}
\end{figure}

In order to examine the effect of such a reduced mass ratio on the
kinetic $\Alf$ wave, we choose once more a propagation angle of $\theta=85^{\circ}$,
and we perform a wavenumber scan for $\beta_{\i}=\beta_{\e}=1$. The
results of this scan are depicted in Fig.~\ref{fig:KAWmass}. As
can be observed in Fig.~\ref{fig:KAWmass}a), the wave frequencies
are not significantly altered by changing the mass ratio, although
small differences are visible close to the ion cyclotron frequency.
On the other hand, the damping rates plotted in Fig.~\ref{fig:KAWmass}b)
show a visible discrepancy (about 65\% overprediction for reduced
mass ratio) across the ion range of wavenumbers, and more significantly
so in the range $1\lesssim kd_{\i}\lesssim4$, where ion Landau damping
is weakened and ion cyclotron damping is not yet active. 

\begin{figure}
\subfloat[]{\includegraphics[width=0.5\textwidth]{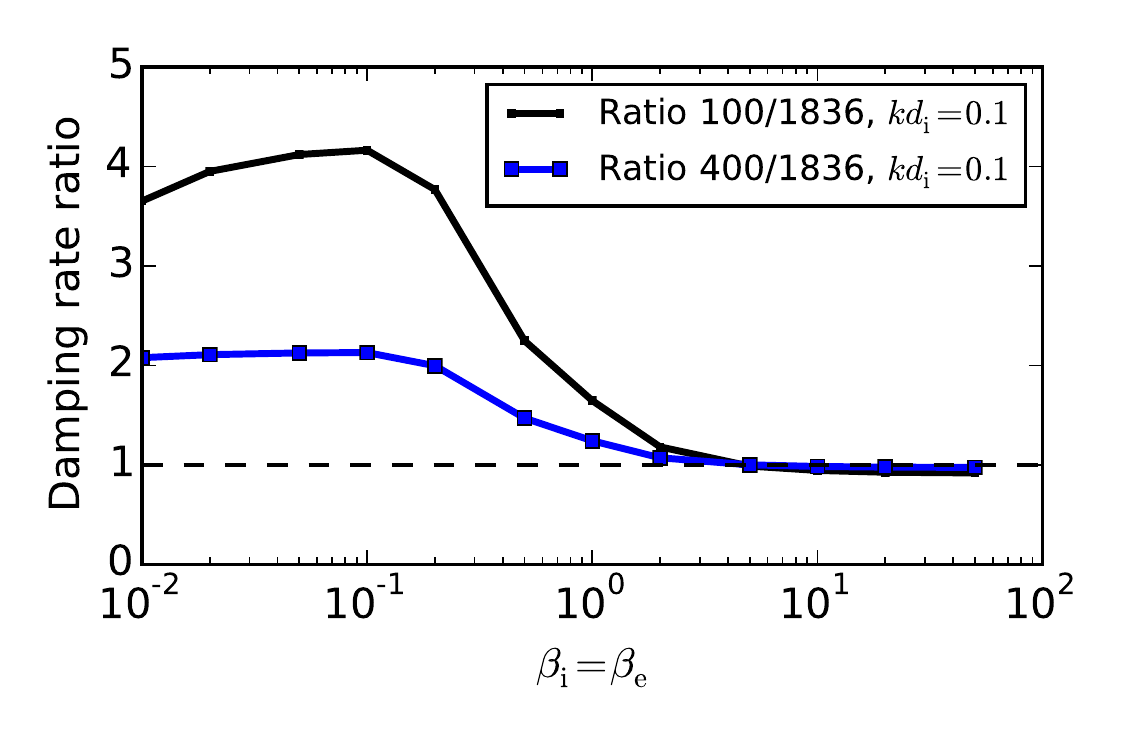}

}\subfloat[]{\includegraphics[width=0.5\textwidth]{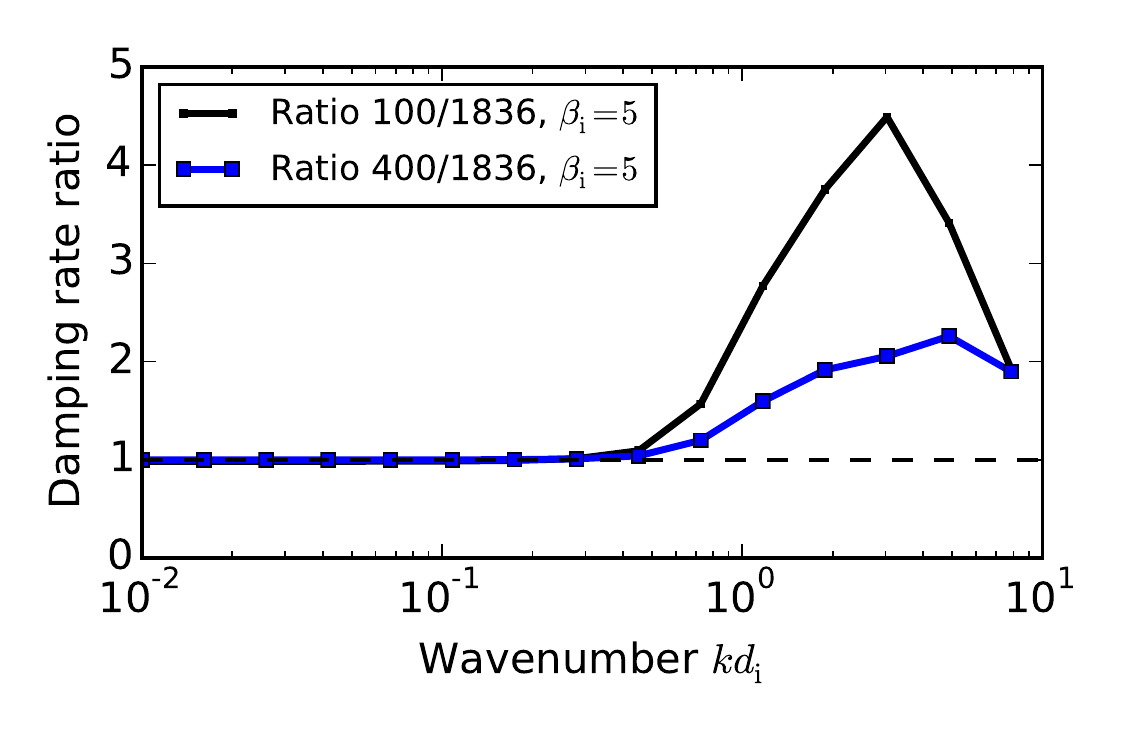}

}

\caption{Ratios between the damping rates obtained with reduced ion/electron
mass ratio and the real proton/electron mass ratio. a) Scan over $\beta$
for $kd_{\protect\i}=0.1$, b) scan over wavenumber for fixed $\beta_{\protect\i}=\beta_{\protect\e}=5$.
Both figures use $\theta=85^{\circ}$. \label{fig:KAWmass-beta}}
\end{figure}

Considering the findings of Sec.~\ref{sub:KAWbeta}, we must suspect
a beta dependence of this finding, and based on the discussion at
the end of that section, we may expect to find more significant differences
at lower $\beta$, where electron Landau damping dominates. In Fig.~\ref{fig:KAWmass-beta}a),
we present the results of a $\beta$ scan, at fixed wavenumber $kd_{\i}=0.1$.
As expected, larger discrepancies (up to a factor 4) occur at low
$\beta$, where electron Landau damping is dominant. 

Another finding is related to the switching of the KAW polarization
that was observed in Fig.~\ref{fig:KAWbeta5} for $\beta=5$. The
exact point in parameter space where this switching occurs depends
on the ion/electron mass ratio, so that different results are obtained
both for frequencies and damping rates in that regime, depending on
the mass ratio employed. In Fig.~\ref{fig:KAWmass-beta}b), another
wavenumber scan is shown to illustrate this for $\beta=5$, where
different mass ratios yield damping rate differences up to a factor
4.5 at large $kd_{\i}$. These discrepancies occur because the wavenumber
where ion cyclotron damping becomes dominant shifts with mass ratio,
exposing the variation in (the otherwise subdominant) electron Landau
damping. 


\subsection{Fast magnetosonic waves/Whistler modes}

In this section, the focus lies on the fast magnetosonic mode and
its potential transition to a whistler mode which exhibits an $\omega\propto k^{2}$
scaling. This wave is ordered out of gyrokinetics, reducing the set
of models to just the hybrid-kinetic and the fully kinetic theory.
The procedure adopted here is similar to that for the kinetic $\Alf$
wave, and we first analyze the dependence of the dispersion relation
on the wavenumber, for two different propagation angles.

\subsubsection{Propagation angle dependence}

\begin{figure}
\subfloat[]{\includegraphics[width=0.5\textwidth]{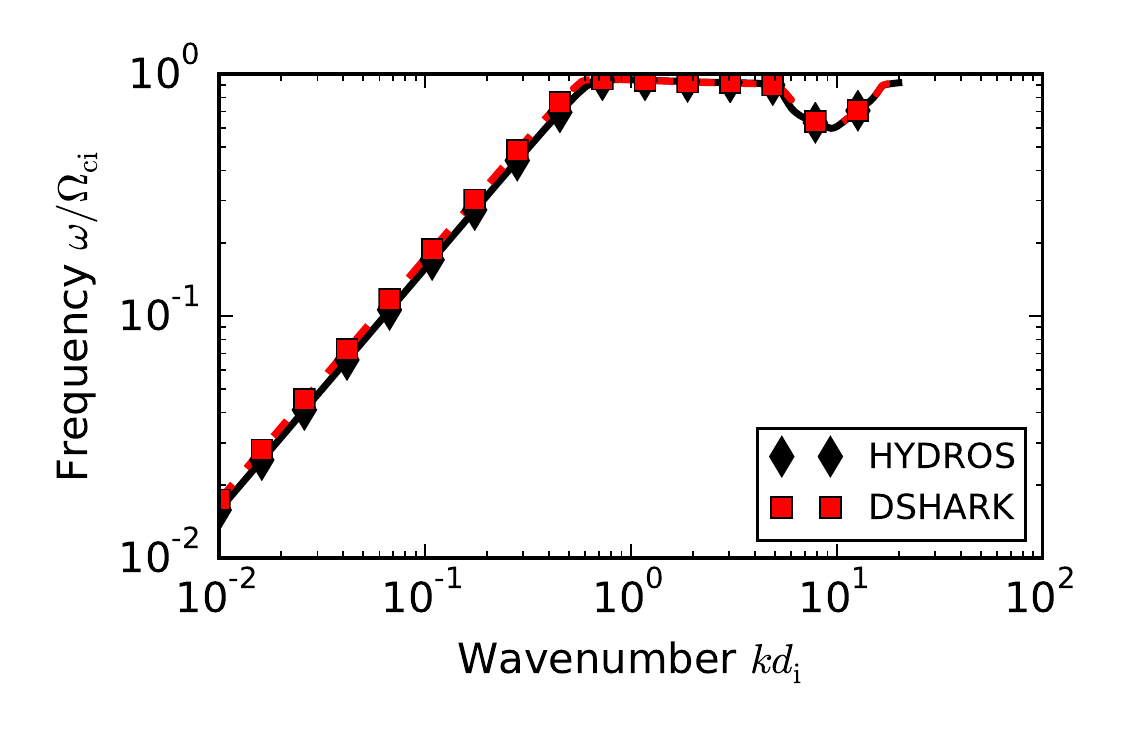}

}\subfloat[]{\includegraphics[width=0.5\textwidth]{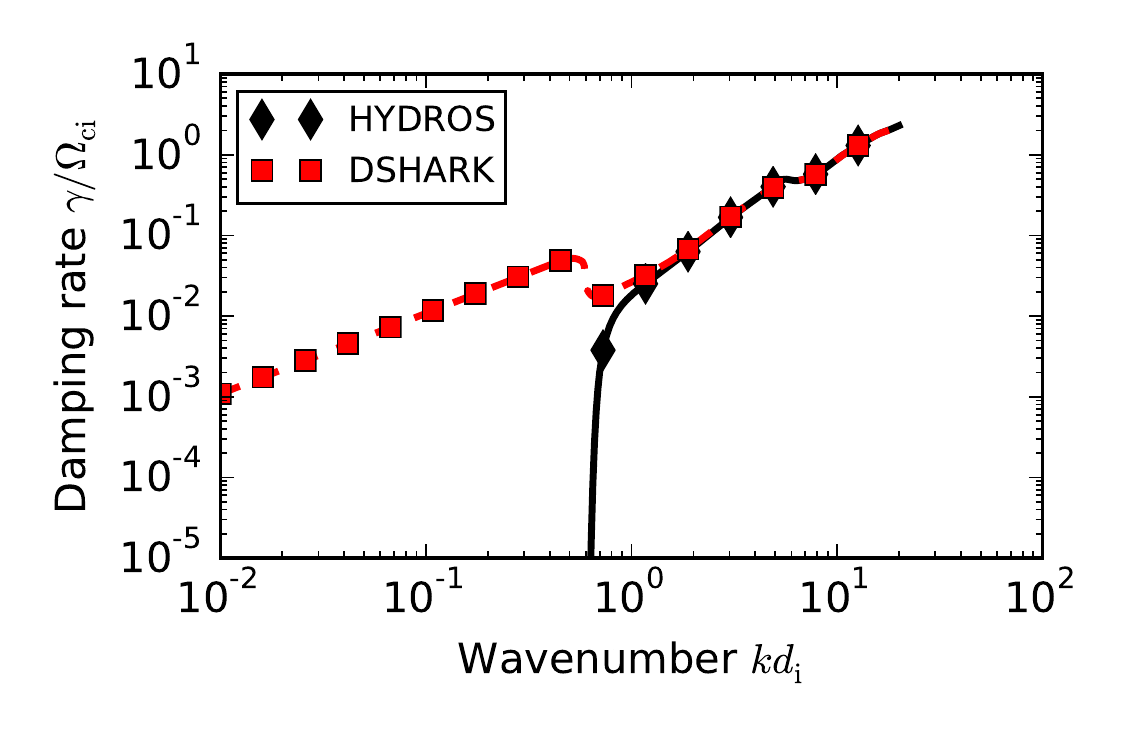}

}

\caption{a) Frequencies and b) damping rates obtained from the HYDROS and DSHARK
solvers for $\beta_{\protect\i}=\beta_{\protect\e}=1$ and a propagation
angle of $\theta=87.5^{\circ}.$ The lines contain all data points,
with every 25th data point emphasized by a marker.\label{fig:Wprop1}}
\end{figure}

As in Sec.~\ref{sub:Kawprop}, a fixed propagation angle of $\theta=87.5^{\circ}$
is chosen (roughly the average propagation angle in the solar wind
plasma) and the wavenumber is scanned for $\beta_{\i}=\beta_{\e}=1$.
The result of this scan is plotted in Fig.~\ref{fig:Wprop1}. Both
models exhibit the same qualitative behavior of the frequency with
a linear $\omega\propto k$ relation until the ion cyclotron frequency
is reached. For the chosen propagation angle, this wave experiences
ion cyclotron damping and stays below the cyclotron frequency. Despite
the qualitative agreement, we note that a shift between the two frequency
curves can be observed in the range $0.01\leq kd_{\i}\lesssim0.6$,
until the ion cyclotron frequency is reached. This deviation was found
before in Ref.~\cite{Told16aarX} and could be traced back to the
massless electron approximation underlying the hybrid-kinetic model.
Conversely, it is possible to obtain the same result from the fully
kinetic dispersion solver by numerically approximating the massless
electron limit. 

In the same wavenumber range, there is a striking difference in the
measured damping rates: while DSHARK reports a rather strongly damped
fast mode ($\left|\gamma\right|/\omega\lesssim0.1$), HYDROS finds
a completely undamped mode (within the numerical accuracy of the code).
With increasing wavenumber, agreement is only obtained when the ion
cyclotron frequency is approached, and for $kd_{\i}\gtrsim1$ the
codes then agree very well on both frequencies and damping rates. 

This severe discrepancy at low wavenumbers (once more, on \emph{ion}
scales) is due to electron transit time damping (i.e. caused by the
near-constant mirror force observed by an electron that travels at
similar speed to the fast wave) \cite{Barnes66}. This effect is not
contained in the hybrid-kinetic model solved by HYDROS, explaining
the observed deviations. In contrast to the kinetic $\Alf$ wave,
this effect plays an important role even for $\beta=1$, which can
be explained by the larger phase velocity of the fast mode compared
to the KAW: in order to obtain a resonant behavior with electrons,
the wave needs to travel at a velocity $v_{\mathrm{ph}}\lesssim\vte$.
The phase velocity of the fast wave for these parameters is roughly
given by $\omega\approx\sqrt{3}\kx\va\approx40\kp\va=40\kp\vti\approx\kp\vte$
(using Eq.~(14) from Ref.~\cite{Gary86}), in good agreement with
the above condition. Interestingly, when the wave reaches the cyclotron
frequency, its linear frequency dependence on $k$ is lost, leading
to a detuning of the resonance and a relatively sharp reduction of
the damping rate, above $kd_{\i}\approx0.6$.

\begin{figure}
\subfloat[]{\includegraphics[width=0.5\textwidth]{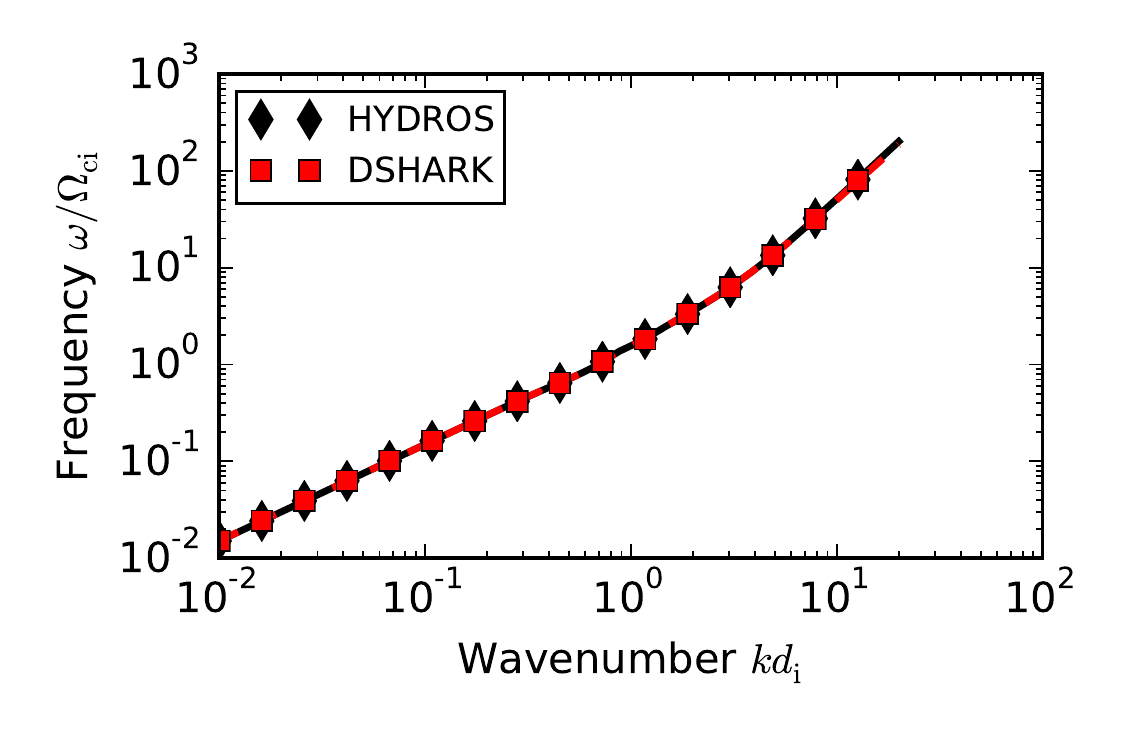}

}\subfloat[]{\includegraphics[width=0.5\textwidth]{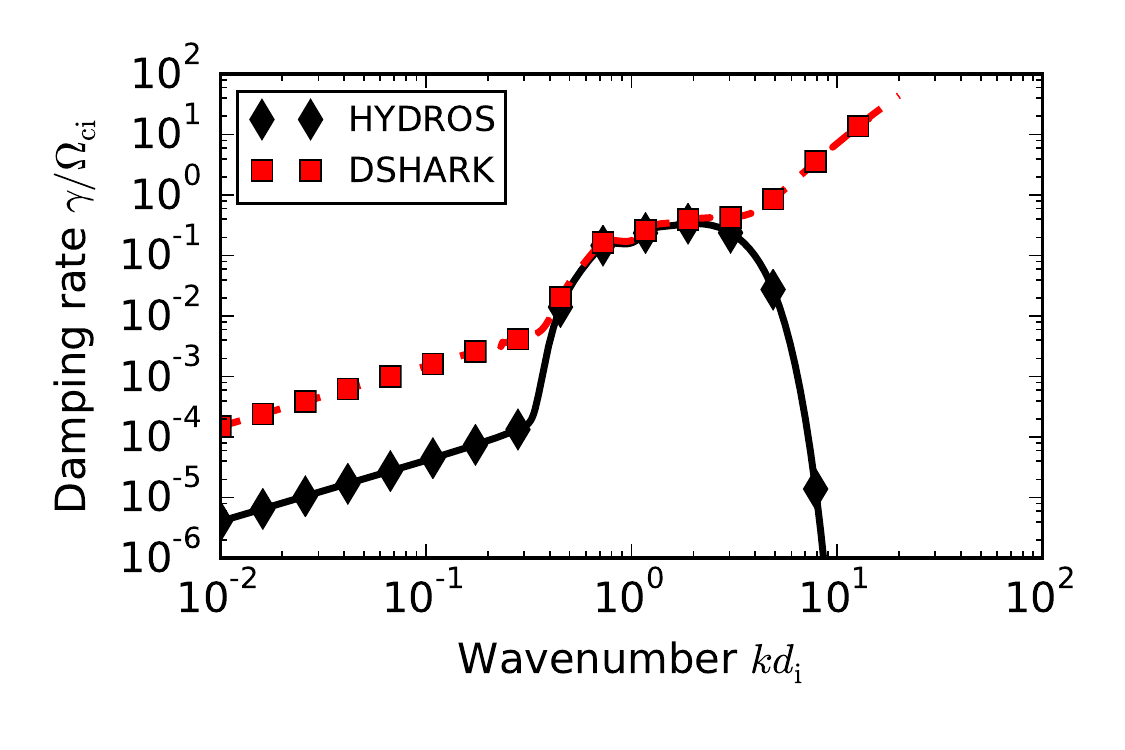}

}

\caption{a) Frequencies and b) damping rates obtained from the HYDROS and DSHARK
solvers for $\beta_{\protect\i}=\beta_{\protect\e}=1$ and a propagation
angle of $\theta=60^{\circ}.$ The lines contain all data points,
with every 25th data point emphasized by a marker.\label{fig:Wprop2}}
\end{figure}

Next, let us study the fast magnetosonic mode/whistler dispersion
for a propagation angle of $60^{\circ}$, keeping $\beta_{\i}=\beta_{\e}=1$.
The results of these scans are presented in Fig.~\ref{fig:Wprop2}.
For these parameters, very good quantitative agreement of the wave
frequencies between the hybrid-kinetic and fully kinetic solvers is
found. In addition, the fast wave now passes smoothly through the
ion cyclotron resonance, and for wavenumbers close to $kd_{\i}\approx3$,
the dispersion relation transitions from its linear (in $k$) to the
well-known quadratic whistler behavior. Although not shown in more
detail here, this behavior of the wave is found (for $\beta=1$) for
propagation angles $\theta\lesssim63^{\circ}$. For more oblique angles,
the fast magnetosonic wave experiences ion cyclotron damping and ceases
to propagate at the ion cyclotron frequency. Note that for such parameters
high-$k$, high-frequency waves with an $\omega\propto k^{2}$ dependence
are still found, but they do not smoothly connect to the sub-$\oci$
fast magnetosonic mode.

Examining in turn the linear damping rates, again a region is found
where electron transit time damping dominates ($kd_{\i}\lesssim0.4$),
leading to a two order of magnitude discrepancy between HYDROS and
DSHARK in that regime. At higher wavenumbers, even though the wave
does not have a cut-off at $\oci$, it still experiences enhanced
damping rates between $0.4\lesssim kd_{\i}\lesssim3$. This damping
is ion cyclotron damping and is thus captured well by HYDROS. At still
higher wavenumbers, electron damping dominates again, widening the
gap between the hybrid-kinetic and fully kinetic damping rates again.
Finally, we note that, although not shown here, for even less oblique
propagation (i.e. moving further away from observed solar wind propagation
angles) ion Landau damping becomes increasingly important, thus improving
the agreement between the two models significantly. 

\subsubsection{Beta dependence}

Following the strong discrepancy between completely undamped (hybrid-kinetic)
and rather heavily damped (fully kinetic) fast modes in the previous
section, we will now characterize the parameter space for which this
finding is of relevance. Indeed, motivated by the findings regarding
the beta dependence of electron Landau damping of kinetic $\Alf$
waves, the same kind of analysis is now performed for the fast magnetosonic
mode, as similar arguments may be expected to hold for this kind of
wave. 

\begin{figure}
\subfloat[]{\includegraphics[width=0.5\textwidth]{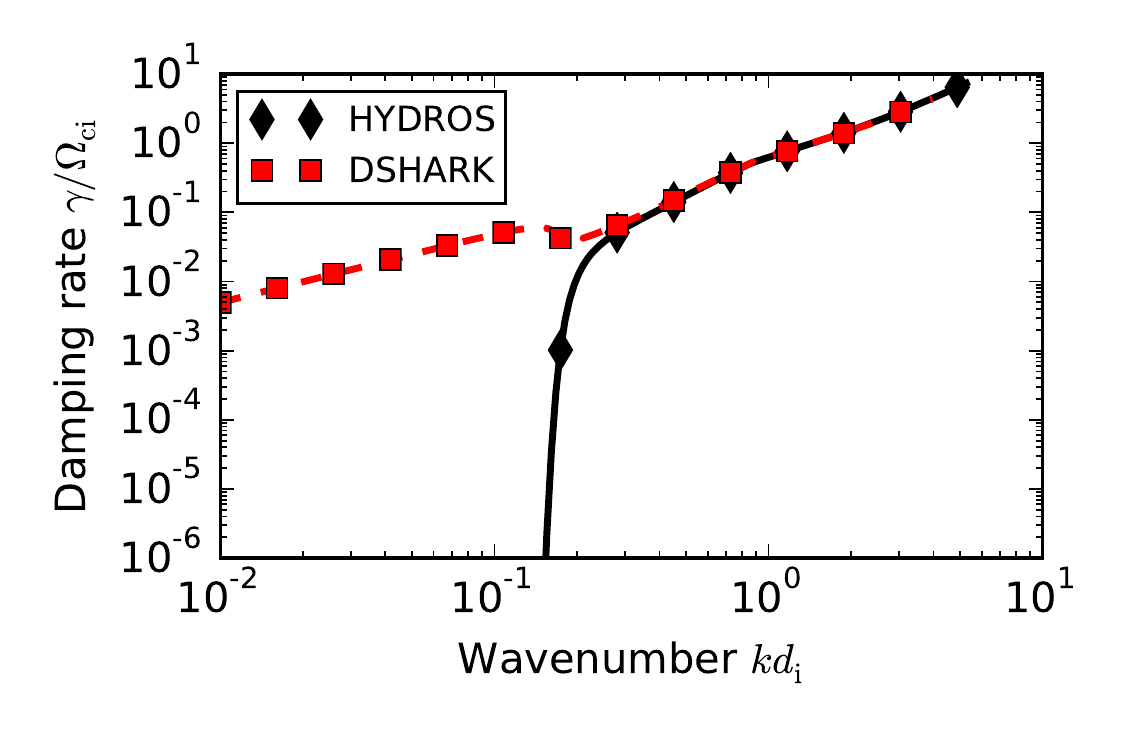}

}\subfloat[]{\includegraphics[width=0.5\textwidth]{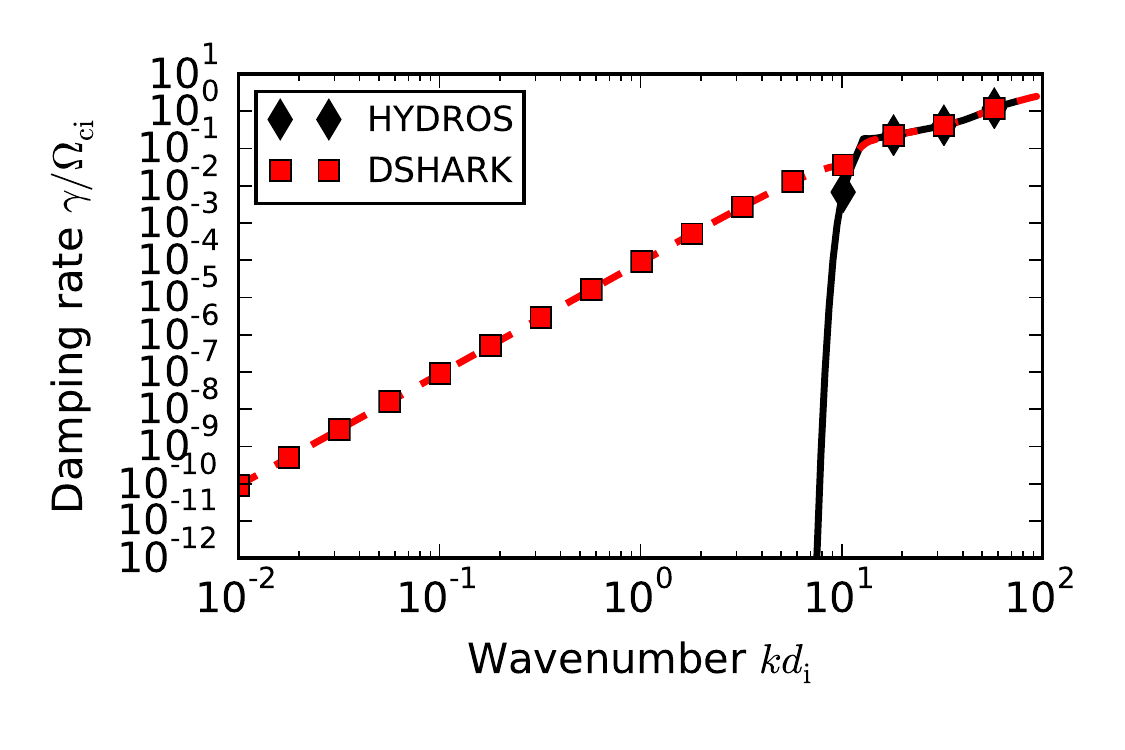}

}

\caption{Fast magnetosonic mode damping rates for a) $\beta_{\protect\i}=\beta_{\protect\e}=20$
and b) $\beta_{\protect\i}=\beta_{\protect\e}=0.01$, for a propagation
angle of $\theta=85^{\circ}$.\label{fig:Wbeta}}
\end{figure}

As before in Sec.~\ref{sub:KAWbeta}, the propagation angle is fixed
to $\theta=85^{\circ}$ and wavenumber scans are performed for various
values of $\beta_{\i}=\beta_{\e}$. In Fig.~\ref{fig:Wbeta}, only
damping rates are shown, since the frequencies generally exhibit satisfactory
agreement between the hybrid-kinetic and fully kinetic models. In
Fig.~\ref{fig:Wbeta}a), a high beta example for $\beta=20$ is chosen,
remembering that for KAWs ion Landau damping was found to dominate
in that parameter regime, masking the lack of electron physics in
the hybrid model. For the fast magnetosonic mode, however, even for
the quite high $\beta=20$ there is a significant wavenumber region
up to $kd_{\i}\approx0.15$ where the wave is undamped in the hybrid-kinetic
model, in contrast to the fairly strong damping ($\left|\gamma\right|/\omega\approx0.1$)
that occurs in the fully kinetic system. At larger wavenumber, cyclotron
damping dominates, which is in turn well matched by the hybrid-kinetic
model. 

This picture remains qualitatively the same for a wide range of $\beta$
values, although the curves shift (in $d_{\i}$ normalization) in
wavenumber space. Fig.~\ref{fig:Wbeta}b) depicts a similar wavenumber
scan for a low $\beta=0.01$. Here, because of the lower wave frequency
of the fast magnetosonic mode at low $kd_{\i}$ a very wide undamped
wavenumber range is found in the hybrid-kinetic model, with cyclotron
damping occurring only at wavenumbers of $kd_{\i}\gtrsim10$. The
mode is more weakly damped at this low $\beta$ value, but the qualitative
picture from before remains intact. 

The very wide $\beta$ range across which electron transit time damping
is found to dominate the picture here can be explained, as was hinted
before, by the much wider electron velocity distribution, owing to
their large thermal velocity. Heuristically speaking, when varying
$\beta$, a wave is detuned much more easily from an ion (Landau or
transit-time) resonance than from an electron resonance. 

\subsubsection{Mass ratio effect}

Finally, we now analyze the effect of a reduced mass ratio on the
fast magnetosonic mode, using the DSHARK solver. The findings of the
previous section, where a massless electron description was found
to result in practically undamped fast modes, suggest that a mass
ratio reduction may have a similarly strong impact. Here, once more
wavenumber scans for the mass ratios $m_{\i}/m_{\e}=1836,$ 400, and
100 are performed, for $\beta_{\i}=\beta_{\e}=1$ and a propagation
angle of $85^{\circ}$.

\begin{figure}
\subfloat[]{\includegraphics[width=0.5\textwidth]{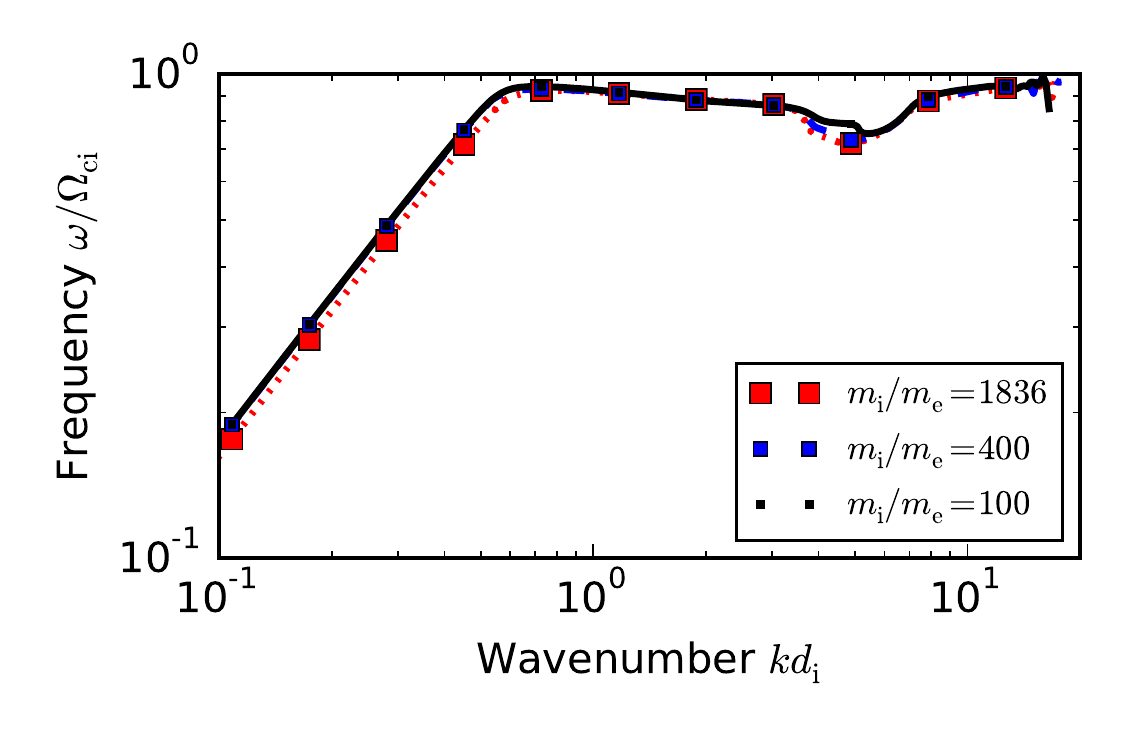}

}\subfloat[]{\includegraphics[width=0.5\textwidth]{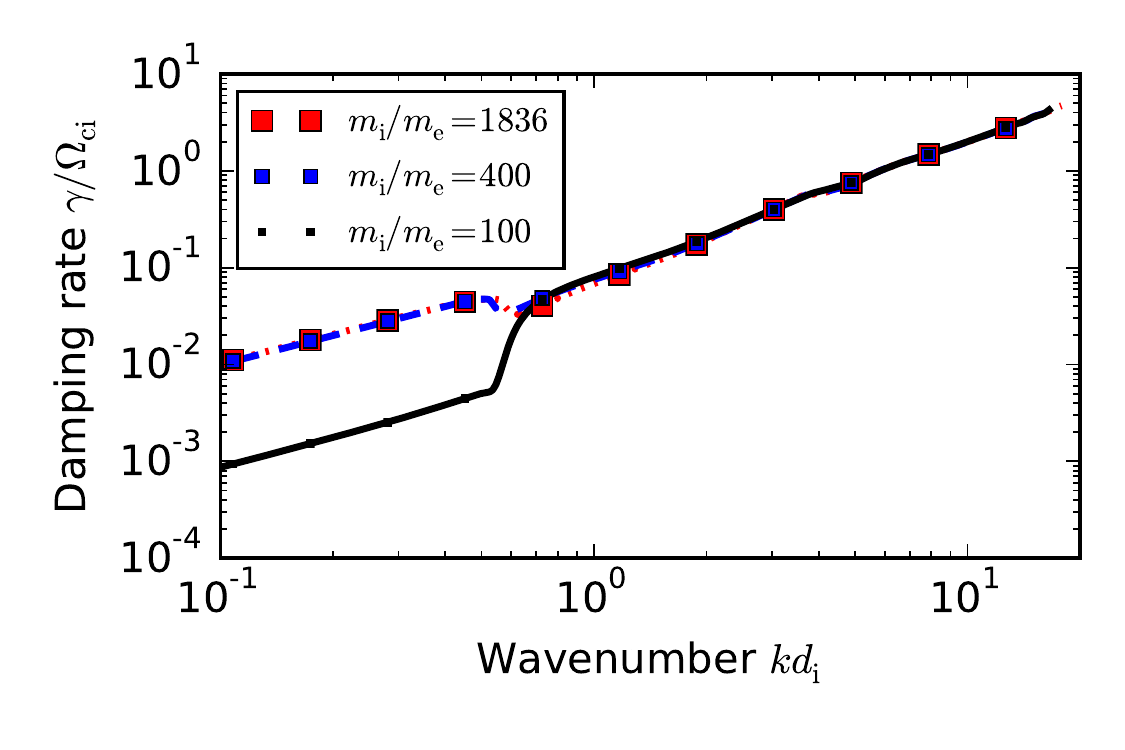}

}

\caption{a) Wave frequency and b) damping rates of the fast magnetosonic mode,
for real proton/electron and reduced mass ratio, for $\beta_{\protect\i}=\beta_{\protect\e}=1$,
and a propagation angle of $85^{\circ}$.\label{fig:Wmass-k}}
\end{figure}

For these parameters, similar to the findings for the KAW, there is
no strong impact of the mass ratio on the wave frequency, see Fig.~\ref{fig:Wmass-k}a).
However, on ion scales below $kd_{\i}\approx0.6$ the wave damping
rate (Fig.~\ref{fig:Wmass-k}b)) turns out to be sensitive to the
mass ratio: while for a mass ratio of $m_{\i}/m_{\e}=400$ the damping
rates deviate only be a few percent, the simulations with a mass ratio
of 100 underestimate the real damping rates by a full order of magnitude.
This behavior can again be understood in terms of the detuning of
the resonance as the electrons are made heavier and heavier -- for
a mass ratio of 100, the fast magnetosonic mode can interact only
with the weakly populated tail of the heavy electron distribution.
At wavenumbers of $kd_{\i}\gtrsim0.6$ on the other hand, ion cyclotron
damping starts to dominate, which is not affected by the reduced mass
ratio. 

\begin{figure}
\subfloat[]{\includegraphics[width=0.5\textwidth]{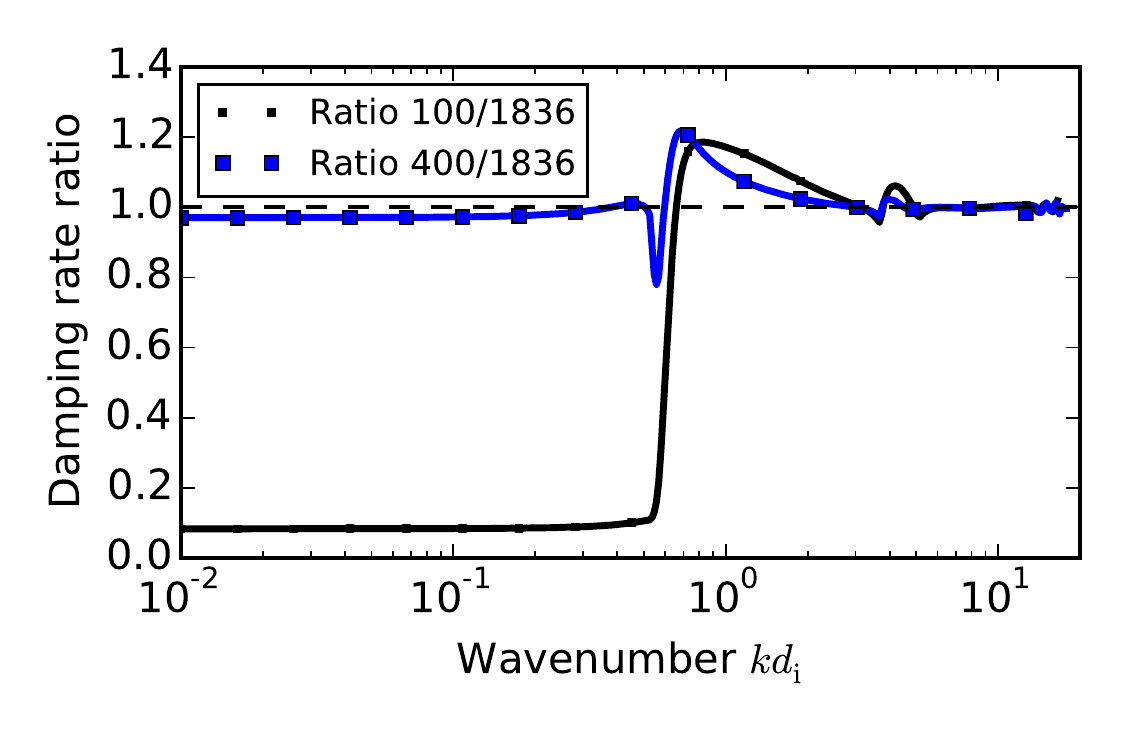}

}\subfloat[]{\includegraphics[width=0.5\textwidth]{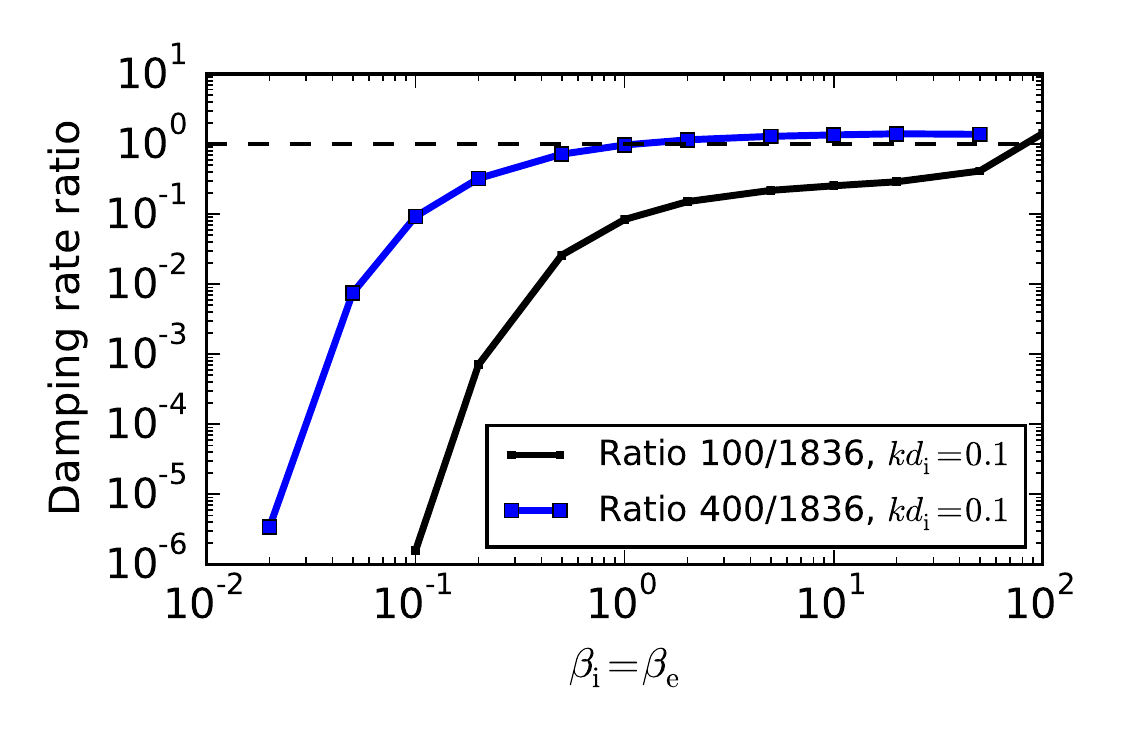}}

\caption{Ratios between the fast wave damping rates obtained with reduced mass
ratios $m_{\protect\i}/m_{\protect\e}=100$, 400 and the real proton/electron
mass ratio. a) Scan over wavenumber for fixed $\beta_{\protect\i}=\beta_{\protect\e}=1$.
b) Scan over $\beta$ for $kd_{\protect\i}=0.1$. Both figures use
$\theta=85^{\circ}$. \label{fig:Wmass-beta}}
\end{figure}

In Fig.~\ref{fig:Wmass-beta}a), the ratios of the damping rates
obtained for reduced mass ratio to the real damping rates are plotted.
As can be seen, a mass ratio of 400 gives a reasonable approximation
to the real damping rates, except in the region where the transition
between the two different damping mechanisms occurs. In contrast to
that, for a mass ratio of 100 the real damping rate is severely underpredicted
for low wavenumber and is only recovered with reasonable accuracy
in the range of dominant ion cyclotron damping, at large wavenumber. 

Finally, in Fig.~\ref{fig:Wmass-beta}b), these damping rate ratios
are plotted for fixed wavenumber $kd_{\i}=0.1,$ while varying beta
over several orders of magnitude. In simulations with a mass ratio
of $m_{\i}/m_{\e}=400,$ the fast wave damping rates are relatively
well represented for large beta (with a discrepancy of about 30\%
for $\beta\gg1$), but reach a discrepancy of more than an order of
magnitude when $\beta<0.1$. The result for $m_{\i}/m_{\e}=100$ is
more concerning, however, as over the entire range of examined beta
values the damping rates are underestimated by at least a factor 2.5,
and by more than an order of magnitude for $\beta<1$. The only exception
occurs at $\beta=100$, where for the fixed wavenumber of $kd_{\i}=0.1$
the fast wave reaches a frequency close to $\oci$, so that ion cyclotron
damping dominates here, which does not depend on the mass ratio. 

\section{Summary and conclusions}

In the present study, the linear wave physics contained in the gyrokinetic,
the hybrid-kinetic (combining a fully kinetic ion treatment and fluid
electron model) and the full kinetic model of plasma physics were
analyzed and compared. For this purpose, we focused on two different
wave modes, namely the kinetic $\Alf$ wave and the fast magnetosonic/whistler
wave. For both kinds of wave the dispersion relations were studied
for two different propagation angles, one of them chosen to be 87.5$^{\circ}$,
close to the average propagation angle found in solar wind plasmas,
and the other close to the maximal observed deviation from the aforementioned
average angle, namely to $60^{\circ}$ \cite{Narita11}.

For the kinetic $\Alf$ wave (KAW), it was found that the gyrokinetic
model (GK) generally agrees very well with full kinetics as long as
the ion cyclotron frequency $\oci$ is not reached. In some cases,
this is even true for frequencies much higher than $\oci$, provided
that the KAW is right-hand polarized. However, energy transfer between
KAWs and other waves present at such high frequencies (that are missing
from GK) cannot be accounted for within the gyrokinetic model. 

As expected, the hybrid-kinetic model was found to agree very well
with the fully kinetic model, as long as electron wave-particle interactions
do not dominate the wave damping. If they do, however, the hybrid-kinetic
model often severely underestimates the linear wave damping rates,
and, perhaps counterintuitively so, even on \emph{ion} spatial scales.
Since the phase velocity of the KAW is close to the $\Alf$ velocity
(and $\va\propto\vti/\sqrt{\beta_{\i}}$), the KAW detunes from the
ion Landau resonance at low $\beta$. Then, electron Landau damping
is the dominant damping mechanism, resulting in an underprediction
of damping rates by the hybrid-kinetic model. 

The fast magnetosonic wave, on the other hand, is ordered out from
GK (even in cases where its frequency is below $\oci$), so only the
hybrid-kinetic and fully kinetic models could be compared here. For
a propagation angle of 87.5$^{\circ}$ and $\beta_{\i}=\beta_{\e}$=1,
a striking disagreement between the hybrid-kinetic and the fully kinetic
model is found: while the wave is rather strongly damped in the latter
case, the hybrid model finds a completely undamped mode. In this case,
the discrepancy is caused by the lack of electron transit time damping
in the hybrid model. Even for less oblique propagation ($\theta=60^{\circ}$),
the damping rates from both models still disagree by almost two orders
of magnitude on ion scales, although ion Landau damping then starts
to become more important. The beta dependence of this effect was studied,
showing that this observation is robust across a wide range of beta
values, owing to the broad (in velocity space) electron transit time
resonance. 

Motivated by the fact that many fully kinetic simulations are carried
out with reduced mass ratio in order to make these simulations feasible,
the fully kinetic model with reduced mass ratio was benchmarked against
its real-mass ratio counterpart. Using a mass ratio of $m_{\i}/m_{\e}=100$,
the effect on KAWs was relatively moderate and mostly limited to $\beta\lesssim1$.
The fast magnetosonic mode, on the other hand, was more gravely affected
and it was found that ion scale damping rates of these modes are underestimated
by at least a factor 2.5 across the whole studied beta range, and
up to several orders of magnitude for $\beta\ll1$. While 3D fully
kinetic simulations using the real proton/electron mass ratio will
remain very demanding for the time being, the findings of this work
should be of use for interpreting existing and future simulations,
and for obtaining projections to the real systems they are meant to
describe. 

Finally, we would like to remark that the studies performed in this
work have barely scratched the surface of the comparisons that are
possible. Given the popularity of both gyrokinetic and hybrid-kinetic
simulations, we believe that the availability of easy-to-use dispersion
solvers for both these models is essential for a realistic assessment
of the models and the simulations performed with them. For this reason
the hybrid-kinetic solver HYDROS has been made available on Github
\cite{Hydros}, as was the DSHARK solver before \cite{DSHARK}.

\label{sec:conclusions}

\section*{Acknowledgments}

D.~T. is grateful to V.~Bratanov for pointing out a mathematical
subtlety in the gyrokinetic dispersion relation. J.~C. was supported
by the NSF REU Grant No. PHY-1460055. The research leading to these
results has received funding from the European Research Council under
the European Union's Seventh Framework Programme (FP7/2007-2013)/ERC
Grant Agreement No. 277870. Furthermore, this work was facilitated
by the Plasma Science and Technology Institute at the University California,
Los Angeles, and by the Max-Planck/Princeton Center for Plasma Physics. 

\bibliographystyle{unsrt}

\end{document}